\documentclass[aps,jcp,twocolumn,groupedaddress,showpacs]{revtex4}
\usepackage{graphicx,amsmath,array}
\begin{document}

\title{Thermoreversible Associating Polymer Networks: I. Interplay of Thermodynamics, Chemical Kinetics, and Polymer Physics}
\author{Robert S. Hoy$^{1}$}
\email{robert.hoy@yale.edu}
\author{Glenn H. Fredrickson$^{1,2}$}
\affiliation{Materials Research Laboratory$^{1}$ and Departments of Chemical Engineering and Materials$^{2}$ , University of California, Santa Barbara, CA 93106}

\date{November 4, 2009}

\begin{abstract}
Hybrid molecular dynamics/Monte Carlo simulations are used to study melts of unentangled, thermoreversibly associating supramolecular polymers. 
In this first of a series of papers, we describe and validate a model that is effective in separating the effects of thermodynamics and chemical kinetics on the dynamics and mechanics of these systems, and is extensible to arbitrarily nonequilibrium situations and nonlinear mechanical properties.
We examine the model's quiescent (and heterogeneous) dynamics, nonequilibrium chemical dynamics, and mechanical properties.
Many of our results may be understood in terms of the crossover from diffusion-limited to kinetically-limited sticky bond recombination, which both influences and is influenced by polymer physics, i.\ e.\ the connectivity of the parent chains.
\end{abstract}
\maketitle

\section{Introduction}
\label{sec:intro}

Flexible synthetic polymers have long been of fundamental scientific interest because many of their properties  arise from a few universal features like the topological connectivity, random-walk like structure, and excluded volume of the chain molecules. 
Less universal are the various attractive, ``associative'' intermolecular interactions \cite{rubinstein99,brunsveld01} ranging from weak dispersion forces to strong covalent chemical bonds (in chemically crosslinked systems).
Examples include hydrogen bonding, electrostatic attractions, and effective attractions driven by incompatibility with a solvent.
These interactions lead to formation of supramolecular structures ranging from micelles to network gels.

Associating polymers (APs) differ from simple homopolymers in that chains contain a (typically fairly low) fraction of ``sticky'' monomers, which are different from the majority-species monomers.
The sticky monomers form ``sticky bonds'' with each other via associative interactions weaker than permanent covalent bonds.
These lead to formation of supramolecular aggregrates.
Unlike the closely related ``living'' or ``equilibrium'' polymers, the degree of polymerization of AP ``parent'' chains is fixed (in time) by permanent covalent backbone bonds.

The lifetime of the sticky bonds is finite.
Depending on the nature of the associative interactions and ambient conditions (e.\ g.\ temperature, concentration), the supramolecular topology may be practically permanent, in which case the system forms a ``chemical gel'' (i.\ e.\ a crosslinked rubber), or so short-lived that the system is indistinguishable from a simple polymer solution, melt or glass.
Between these limits, when the topology of the associated supramolecular aggregates changes on a time scale comparable to the experiment, these systems form complex fluids with fascinating dynamical and mechanical properties \cite{rotello08}.

At fixed ambient conditions, the time scales for topological changes in associating polymer systems are in principle set by three independent factors:  the (a) thermodynamics (i.\ e.\ energetic strength relative to $k_B T$), (b) the ``chemical kinetics'' of the sticky bonds, and (c) the underlying non-associative polymer physics.
Thermodynamics set ``static'' quantities such as the size of the supramolecular aggregates and hence the position of the system relative to the percolative gelation transition.
Kinetics set relaxation times through their effect on the rates of formation and breaking of sticky bonds.
Polymer physics alters the dynamics through such effects as the random-walk-like structure and uncrossability of chains, which give rise to the systems' underlying Rouse or reptation dynamics \cite{doi86}.
The interplay of (a)-(c) allows for the design of materials with exquisitely tunable rheological response.
For this reason, APs have been the focus of intense experimental and theoretical study over the past two decades; see Refs.\ \cite{brunsveld01,rotello08,annable93,binder07,tanaka02} for reviews.

Changes in ambient conditions lead to ``thermoreversible'' property changes unique to AP systems, e.\ g.\ extremely sharply decreasing viscosity upon decreasing concentration or increasing temperature.
These changes can be tuned (engineered), so APs have great potential as ``smart'' materials \cite{brunsveld01,binder07,rotello08,deGreef08,cordier08} in which the change of lifetime or concentration of the sticky bonds with ambient conditions leads to useful products.
Applications include temperature-sensitive adhesives, coatings for heat-sensitive materials, and generally enhanced melt processability relative to conventional polymers \cite{deGreef08}.

Static thermodynamic properties of AP systems, in particular the percolative gelation transition and local structural changes arising from associative interactions, have been extensively studied.
Analytic theories provide a good understanding of homogenous systems, and emerging numerical techniques such as self consistent field theoretic simulations \cite{lee07} and reaction-ensemble DPD \cite{lisal09} show promise for investigating inhomogeneous systems.

However, the dynamical, mechanical, and nonequilibrium properties of associating polymer gels and networks remain poorly understood.
Time dependent properties obviously depend on kinetics, and in addition to being of fundamental scientific interest, a better understanding of them may prove important in developing new applications of associating polymer systems such as self healing materials \cite{cordier08}.
The situation becomes particularly complex when the lifetime of the sticky bonds is not long or short compared to the ``polymeric'' relaxation times; this regime has been studied rather extensively for linear equilibrium polymers (see e.\ g.\ Refs.\ \cite{cates87,oshaugh95,combinedhuang01,nyrkova07,stukalin08}), but much less so for networks.

Analytic and quasi-analytic approaches to AP dynamics and mechanics, e.\ g.\ \cite{foot7} Refs.\ \cite{green46,cates88,baxandall89,leibler91,tanaka92,groot96,rubinstein98,vaccaro00,rubinstein01,tanaka02,semenov06,indei07,semenov07,tripathi07}, have made many useful, experimentally verifiable predictions, including nonlinear behaviors such as shear thickening and strain hardening \cite{pellens04,tripathi07}.
In the general case, however, the complex interplay of sticky bond thermodynamics and kinetics with the underlying polymer physics in these systems is almost certainly beyond the reach of analytic theory.
For the sake of tractability, theories have generally neglected one or more features of AP systems that are likely essential to capturing their behavior under certain ambient conditions.
For example, as temperature drops towards the glass transition, attractive, non-associative interactions, such as van der Waals forces between non-sticky monomers, become increasingly important \cite{baschnagel05}.
Moreover, virtually all analytic treatments have thus far been restricted \cite{foot9} to homogeneous AP systems; the majority focus on the ``telechelic'' case of APs with only 2 sticky monomers per parent chain (one on each end). 
We believe that inhomogeneous AP systems are the potentially the most interesting and useful, e.\ g.\ because inhomogeneities serve to localize sticky monomer concentration and network connectivity, which in turn can broaden the relaxation spectrum \cite{kumar07,katieunpublished}.

The above set of potentially essential features of APs is not treated microscopically by existing theories, but can be readily captured by particle-based simulations.
This is the first of a series of simulation studies, the goal of which is to elucidate the separate effects of sticky bond thermodynamics, kinetics, and other underlying polymer physics on the dynamical, mechanical, and nonequilibrium properties of associating polymers.

Previous particle based simulations of AP networks focusing on dynamical properties, e.\ g.\ \cite{foot7} Refs.\  \cite{groot94,nguyen95,liu96,milchev99,kumar01,ayyagari04,guo05,manassero05, loverde05,combinedloverso,baljon07,goswami07} have produced a wealth of interesting ideas and results.
However, for various reasons outlined in Section \ref{subsec:compareprevmethods}, the methods employed in these previous studies are not suitable for the full range of problems we wish to consider.
A more versatile model for APs requires the combination of: \textbf{i)} realistic dynamics, \textbf{ii)} applicability to far-from-equilibrium conditions,  \textbf{iii)} controllable sticky binding topology, \textbf{iv)} variable chemical kinetics and \textbf{v)} the ability to treat inhomogeneous systems.

In this paper we develop and validate a model, based on those employed in Refs.\ \cite{kremer90, combinedhuang01,baljon07}, that shows all these characteristics.
Much of the physics of amorphous polymer melts and gels is independent of chemical detail \cite{doi86,rubinstein03}, so properties i) and ii) are captured by using a bead-spring model \cite{kremer90} for the underlying (non-associative) polymer physics.
We capture properties (iii)-(v) by employing a hybrid molecular dynamics / Monte Carlo (MD/MC) approach with controlled (specifically, binary) bonding and variable chemical kinetics \cite{combinedhuang01}. 
The combination of properties (i-v) allows the model to be used to obtain many results unattainable with previous methods. 
A key advance is that our algorithm is parallelizable.
In validating and investigating our model, we separate the effects of thermodynamics, kinetics, and polymer physics on AP network dynamics and mechanics.
We show that sticky bonding is mappable to a ``mean-field'' two-state Arrhenius rate model, but that the kinetic rate constants for SB association and dissociation are affected by the fact that the SMs are embedded in polymers.

One of the key aspects of associating polymer networks, which has rarely been examined (for \textit{networks}) in the theoretical or simulation literature, is that sticky bond relaxation can be either kinetically limited (i.\ e. limited by the intrinsic rate of bonding and/or debonding) or diffusion limited (i. e. when the kinetics are so fast that sticky bond scission and recombination events become correlated because newly broken SM pairs tend to recombine before exploring the network cage).
Most experimental systems seem to be kinetically limited, and this is the case treated by almost all published analytic theories for reversibly associating networks, including transient network models \cite{tanaka92,vaccaro00}.
Theories for kinetically limited AP networks \cite{baxandall89,leibler91,rubinstein98,rubinstein01,semenov06} assume that the sticky bond lifetime $\tau_{sb}$ is so long compared to all ``polymeric'' relaxation times that it controls all important time scales for network relaxation, and therefore that kinetics only affect key network relaxation times through their effect on $\tau_{sb}$ (or, alternately, as suggested by recent experiments \cite{yount05,loveless05}, the inverse dissociation rate constant $k_b^{-1}$, defined below).
We show that the validity of this assumpton is questionable, and that there is a wide parameter space, plausibly accessible in experiments, where it is invalid.
Note that published theories \cite{degennes82,oshaugh87,oshaugh88,oshaugh95,oshaugh98,fredrickson96} for diffusion-limited reactions in polymeric systems, such as those of O'Shaughnessy \textit{et.\ al.}, either are not immdeiately applicable or have not yet been applied to reversible networks.

Our simulations explore the crossover between the kinetically limited and the diffusion limited cases.
We confirm a key prediction of Rubinstein and Semenov \cite{rubinstein98} on the role of bond recombination on AP dynamics, specifically that recombination effectively renormalizes the SB lifetime in  systems where the sticky bonds are sufficiently strong.
However, we show that recombination couples interestingly to the diffusive-kinetic crossover in a way not previously predicted. 

We extensively examine the dynamics in quiescent, equilibrium systems.
One of the more interesting results is that slowing chemical kinetics increases dynamical heterogeneity \cite{kumar01} in a manner similar to increasing the thermodynamic strength of the sticky bonds.
We also examine nonequilibrium `chemical dynamics' (i.\ e.\ in systems where the initial sticky bond population is not equilibrated), and nonlinear mechanical properties.
The studies of nonequilibrium systems and mechanical properties presented here are limited in number because this paper is intended primarily to illustrate the broad utility of the model.

The organization of the rest of this paper is as follows.
In Section \ref{sec:modelmethods} we further motivate, describe and extensively validate our model.  
We also discuss how it differs from those used in previous simulations of APs.
In Section \ref{sec:results} we present various results for the equilibrium and nonequilibrium physics of thermoreversible AP networks, and compare to theoretical predictions.
Finally, Section \ref{sec:conclude} presents a discussion and conclusions.
Two Appendices include technical details of the analyses.

\section{Models and Methods}
\label{sec:modelmethods}

Of particular interest for the current study, and motivating our modelling approach, are recent experiments performed by the group of Stephen L.\ Craig \cite{yount03,yount05,loveless05,rotello08,xu09,craigexpl}.
These have attempted to \textit{independently} vary thermodynamics and chemical kinetics by making systematic changes in sticky monomer chemistry (based on metal-ligand interactions).
Systems with similar static properties show dramatic differences in time-dependent properties that are directly associated with the different kinetics. 
This effect is quite challenging to capture experimentally because thermodynamics and kinetics are highly correlated for most AP systems (i.\ e.\ stronger binding $\leftrightarrow$ slower kinetics \cite{rotello08}), but relatively easy to impose in simulations.

The advantages of including variable kinetics in the model are rather obvious given the above discussion.
The advantage of imposing binary bonding (i.\ e.\ a SM is at all times bonded to either 0 or 1 other SM) is also relevance to current experiments.
Sticky monomers with binary bonding are considered particularly valuable for making thermoplastic elastomers with enhanced melt processability and controllable network architecture \cite{sijbesma97,sontjens08}.
Real examples include multiple hydrogen bonding monomers such as ureidopyrimidinone (UPy), which have highly directional associative interactions, and are of a strength such that the sticky bonds they form constitute a ``reversible alternative for the covalent bond'' \cite{binder07,deGreef08,sijbesma97,sontjens00,sontjens08}.
In the model used here \cite{baljon07}, sticky bonds differ from covalent bonds only by their reversibility and strength. 

\subsection{Hybrid MD/MC Simulation Protocol}
\label{subsec:descrip}

Our model is built on the framework of the Kremer-Grest bead-spring model \cite{kremer90}, which has been extensively validated and is known to capture the key physics of linear homopolymer melts \cite{kremer90,putz00} as well as permanently crosslinked networks \cite{combinedgrest90}.
Each associating polymer chain is linear and contains $N$ beads (monomers). 
Systems consist of $N_{ch}$ chains, so the total number of monomers is $N_{ch}N$.
Periodic boundary conditions are imposed in all three directions, with periods $L_i$ along directions $i=x$, $y$, and $z$.
Values of $N_{ch}$ range from 700 to 5600; the lowest values are chosen to satisfy $L_i > 2R_{ee}$ (where $R_{ee}$ is the mean chain end-end distance), preventing self interactions.
Remaining leading order finite size effects in networks scale as $(N_{ch}N)^{-1/3}$ \cite{hentschke05} and should be $\sim 2\%$ for the systems considered here.

All monomers have mass $m$ and interact via the truncated and shifted Lennard-Jones (LJ) potential $U_{LJ}(r) = 4u_{0}[(a/r)^{12} - (a/r)^{6} - (a/r_{c})^{12} + (a/r_c)^{6}]$, where $r_{c}$ is the cutoff radius and $U_{LJ}(r) = 0$ for $r > r_{c}$.
Covalent bonds between adjacent monomers on a chain are modeled using the
finitely extensible nonlinear elastic potential $U_{FENE}(r) = -(1/2)(kR_{0}^2) {\rm ln}(1 - (r/R_{0})^{2})$, with the canonical \cite{kremer90} parameter choices $R_{0} = 1.5a$ and $k = 30u_{0}/a^{2}$.   
In this study, following the majority of bead-spring studies on permanently crosslinked systems (e.\ g.\ Refs.\ \cite{combinedgrest90,svaneborg08}), we employ flexible chains with no angular potential.
We express all quantities in units of the LJ bead diameter $a$, intermonomer energy $u_{0}$, and the LJ time $\tau_{LJ} = \sqrt{ma^{2}/u_{0}}$.

All systems have monomer density $\rho = 0.85/a^3$.
We employ two temperatures in this study: $k_B T= 1.0u_0$ and $k_B T= 0.6u_0$.
These ambient conditions both correspond to dense polymer melts far above the glass transition temperature $T_g$; $T_g \simeq 0.35u_0/k_B$ for $r_c = 1.5a$ and decreases with decreasing $r_c$ \cite{rottler03c,bennemann98}.
This far above $T_g$, melt physics is known to be dominated by the repulsive part of the intermonomer interactions \cite{doi86}; for convenience, and following convention \cite{kremer90}, we use purely repulsive LJ interactions with $r_c = 2^{1/6}a$.
However, including attractive interactions by increasing $r_c$ is trivial, is important to realistically capture $T$ dependent properties, and will be done in upcoming studies.

All simulations are performed using an enhanced version of the LAMMPS \cite{plimpton95} MD code.
Newton's equations of motion are integrated with MD using the velocity Verlet method \cite{frenkel02} and typical timestep $\delta t = .01\tau_{LJ}$ \cite{foot1}.
A Langevin thermostat \cite{schneider78} is used to maintain the temperature.
The damping time $\tau_{Lang} = 10-100 \tau_{LJ}$ is larger than the value typically used ($\tau_{Lang} \simeq \tau_{LJ}$) in bead spring studies; this reduces undesirable thermostat-driven effects such as alteration of stress relaxation by suppression of hydrodynamic momentum transfer \cite{pastorino07}.
In this study we employ two ``chain lengths''. 
Most studies are performed at $N = 50$, which is at or below best estimates of the entanglement length $50 \lesssim N_e \lesssim 85$ \cite{everaers04,hoy09}, so the melts can be fully equilibrated by allowing chains to diffuse several $R_{ee}$ \cite{auhl03}.
$N = 50$ is also a convenient choice of chain length because it has been considered in many previous studies.
To elucidate the effects of underlying polymeric structure on AP physics, we also consider monomeric melts ($N=1$), that reversibly sticky-bond into dimers.

After the melts are equilibrated, we choose a fraction $c_{st}$ of the monomers to be ``sticky''.
For the $N=1$ systems these are chosen randomly.
For $N=50$ systems, SMs are placed uniformly along chains: at both chain ends and also at internal monomers $iN/(Nc_{st} - 1)$, where $i = 1, 2, ..., Nc_{st} - 2$.
In this study we use $c_{st}=0.08$ (which is comparable to typical experimental values, e.\ g.\ \cite{yamauchi03,katieunpublished}, so for $N = 50$ the SMs are the 1st, 17th, 34th, and 50th monomers in each chain.
However, any SM placement can be used, and studies of the effects of altering SM placement at fixed $c_{st}$ are underway; the effects of chemical disorder are known to be significant for stress relaxation \cite{semenov07,katieunpublished}.

\begin{figure}[h]
\includegraphics[width=3.25in]{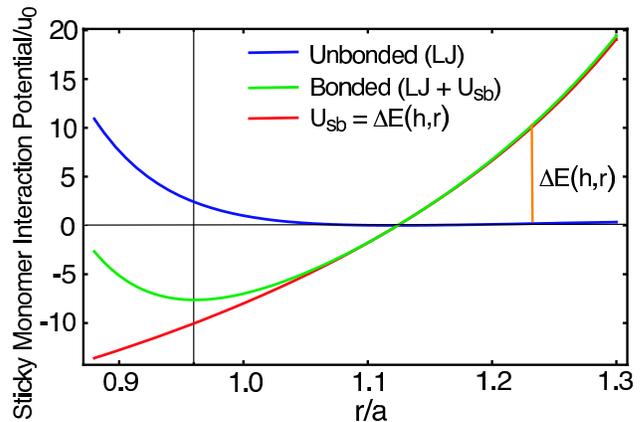}
\caption{Sticky monomer interaction potential.  Same as in Refs.\ \cite{combinedhuang01, baljon07}.  Differences in SB formation/breaking rules are noted in the text.}
\label{fig:smpotential}
\end{figure}

Sticky monomers are identical to regular monomers, except that they form reversible ``sticky'' bonds.
Figure \ref{fig:smpotential} illustrates the potential energy between sticky monomers as a function of their separation $r$.
SMs (like all monomers) always interact via Lennard Jones interactions, whether bonded or not.
Bonded SMs additionally interact via the potential $U_{sb}(r, h)$:
\begin{equation}
U_{sb}(r, h) = U_{FENE}(r) - U_{FENE}(r_0) - h,
\label{eq:Usticky}
\end{equation}
which is based on the standard covalent FENE potential.
Here $r_0$ represents the equilibrium FENE bond length; $r_0 \simeq .96a$, i.\ e.\ the minimum of the potential $U_{LJ}(r) + U_{FENE}(r)$.
The only difference between the sticky and covalent bond potentials is thus an $r$-independent, tunable offset.
The same bonding potential was used in Huang \textit{et. al.}'s studies of equilibrium polymers \cite{combinedhuang01,combinedhuang2} and a very similar potential was used in Baljon \textit{et.\ al.'s} studies of telechelic associating networks \cite{baljon07,foot4}.
However, our method has several important differences from those of Refs. \cite{combinedhuang01,combinedhuang2,baljon07} (see Section \ref{subsec:compareprevmethods}), so we explain it in detail below.

The potential $U_{sb}(r)$ has several other important features.
$h$ represents the sticky binding energy; for $h = 0$ a sticky bond can be formed between two monomers separated by $r_0$ with no change in energy.
The associated force $F_{sb}(r) = -\partial U_{sb}/\partial r$, however, is independent of $h$.
Adjusting $h$ is thus a nearly pure way of adjusting the thermodynamics of the sticky bonds without directly altering their ``chemical kinetics'', i.\ e.\ the rates of formation and dissociation of sticky bonds.

Formation and breaking of sticky bonds is performed using Metropolis Monte Carlo \cite{frenkel02}.
The change in energy required to form a sticky bond between an unbonded pair of SMs is just $\Delta E(r, h) = U_{sb}(r, h)$, and the energy change to break a sticky bond is $\Delta E(r, h) = -U_{sb}(r, h)$.
These are the only MC ``moves'' used, and the acceptance probability of the moves is set by $\Delta E(r, h)/k_BT$.
The MC moves are strictly ``topological''.
All spatial motion of bonded SMs is governed by the sticky bond force $F_{sb}(r)$, along with the other forces from Lennard Jones and covalent FENE interactions, which are all integrated using MD.
One potential difficulty is that only bonded SMs ``feel'' the force $F_{sb}(r)$, so formation/breaking of SMs creates temporal force discontinuities.
However, as will be shown below, this does not seem to cause any spurious behavior.

The system sizes and time scales studied here require simulation on parallel computers; 8 to 64 processors are used in a typical simulation.
While MD parallelizes very well \cite{plimpton95}, it has long been noted that MC \cite{esselink95} is very difficult to parallelize.
We therefore perform hybrid parallel MD / serial MC simulations.
MC moves are performed once every $\tau_0$ in Lennard Jones time units; $\tau_0$ is the MC ``timestep''.
The MD simulation is paused while the parallel-distributed lists of sticky bonds and SM coordinates are gathered onto one processor.
For efficiency, Verlet-style pair neighbor lists of SMs are used and of open SM pairs, only those within $r < R_0$ are considered for SB formation.
After the SB list is updated, it is distributed back to all processors and the MD simulation resumes.
Great care was taken to optimize the MC algorithm to minimize ``dead'' time on the other processors, but reasonable parallel performance requires $\tau_0 \gg \delta t$.
In this paper, except where otherwise noted, we use $\tau_0 = \tau_{LJ}$; see also Section \ref{subsec:twostatedynamics}.

As mentioned above, current experimental trends favor binary-bonding sticky monomers,
We impose binary bonding through a simple restriction on the Monte Carlo routine; sticky bond formation is attempted only for pairs of unbonded SMs.
The 1-1 bonding restriction imposed here was also assumed in Ref.\ \cite{rubinstein98}, which eases comparison of our results to theoretical predictions.

Two further technical details of the MC algorithm are noteworthy:  
(1) We do not allow any SM pair to both break and form a SB (i.\ e., to break and recombine) during the same MC step.
This is a technical violation of detailed balance, but satisfies the weaker ``balance'' condition sufficient \cite{manousiouthakis99} for accurate MC simulations.
SB recombination is a critical aspect of supramolecular polymer physics \cite{rubinstein98} and is further discussed in Section \ref{sec:results}.
(2) we do not allow ``bond switching'' moves within a single MC step.  
That is, for SMs $V,W,X,Y$, we do not allow moves of the form
\begin{equation}
\begin{array}{ccc}
V-W + X & \rightarrow & V-X + W\\
V-W + X-Y & \rightarrow & V-Y + X-W
\end{array}
\label{eq:forbiddenmoves}
\end{equation}
or any other more complicated moves.
In addition to being difficult to implement in simulations, such processes are unlikely to occur instantaneously in real systems, in part because of steric constraints.
Different but analogous rules, suitably modified to the use of SMs with two binding sites each, were imposed in Refs.\ \cite{combinedhuang01,combinedhuang2}. 

In our model, varying the relative rates of sticky bond formation and dissociation is accomplished by varying $h$.
However, the absolute values of the rates depend on a yet unspecified \textit{kinetic} time scale $\tau_{kin}$.
This time can be controlled through the Monte Carlo routine.
At each MC timestep (i.\ e., every $\tau_0$), a fraction $f_{MC}$ of unbonded SM pairs (of those within range $r < R_0$) and an equal fraction $f_{MC}$ of bonded SM pairs are randomly selected to be considered respectively for sticky bond formation and breaking.
We have verified this scheme maintains `balance' \cite{manousiouthakis99} for pairs and triplets of SMs for $.01 \leq f_{MC} \leq 1$.

Thus the average time over which each unbonded or bonded SM pair is considered once for (respectively) SB formation or breaking is  $\tau_{MC} = f_{MC}^{-1}\tau_0$, and the parameter $\tau_{MC}$ effectively controls the ``chemical kinetics'' of the SBs.
For a discussion of why we use $f_{MC} < 1$ rather than varying $\tau_{0}$, see
Section \ref{subsec:twostatedynamics}.
Small $\tau_{MC}$ correspond to fast chemical kinetics \cite{foot16}, while large $\tau_{MC}$ correspond to slow chemical kinetics.

In Section \ref{subsec:nonlinmechprop} we perform mechanical tests on various systems.
Two types of tests are perfomed; constant volume deformation and tensile creep.
In the constant volume deformation tests, $L_z$ is increased at a true strain rate $\dot{\epsilon} = \dot{L_z}/L_z$, and $L_x$ and $L_y$ are adjusted to maintain constant volume.
In the creep tests, a constant (small) stress difference $\sigma_{creep}$ (relative to the equilibrium hydrostatic pressure in the quiescent state, which is positive for repulsive LJ interactions) is applied along the $z$ direction using a Nose Hoover barostat \cite{frenkel02}.  
This smaller $|\sigma_{z}|$ produces tensile creep.  
Both types of tests use $\tau_0 = .2\tau_{LJ}$ to minimize systematic errors.

\subsection{Comparison to Previous Simulation Protocols}
\label{subsec:compareprevmethods}

It is worthwhile to compare the simulation method and ambient conditions described above in the context of previous AP simulation studies.
The use of a hybrid MD/MC method is a powerful advantage.
Pure Monte Carlo (MC) simulations have been performed with lattice \cite{nguyen95, liu96, kumar99, kumar01,wang08} and off-lattice \cite{groot94,loverde05} models.
These are very effective at studying static properties like percolative gelation and (in the case of solutions) phase separation, but have limited ability to capture the complex, collective dynamics which occur in bulk polymers, and thus lack properties (i) and (ii).
For example, MC can not, even in principle \cite{milchev99}, capture hydrodynamic effects, which are expected to play an important role whenever momentum transfer is important (e. g. in relaxation of highly stressed systems).

Pure molecular dynamics (MD) studies \cite{khalatur99,ayyagari04,guo05,manassero05,combinedloverso,goswami07} have been used to study static and dynamic properties. 
While better able to capture dynamics and nonequilibrium phenomena than MC, MD studies can not naturally implement controllable sticky bonding topology.
Also, MD studies cannot easily impose any control of chemical kinetics without resorting to costly, chemically realistic models.
For example, Padding and Boek \cite{padding04} studied systems intermediate between ours and those studied by Huang \textit{et.\ al.}; a fraction $c_{st} < 1$ of their monomers were allowed to form linear equilibrium poiymers, but the FENE-C sticky bonding potential \cite{kroger04} used did not allow for variable kinetics.
Thus, in practice, typical MD studies lack properties (iii) and (iv).

The previous works most closely related to the present method are Refs.\ \cite{baljon07,combinedhuang01,combinedhuang2}, who also used hybrid MC/MD with the same $U_{sb}(r)$ (Eq.\ \ref{eq:Usticky}).  
Huang \textit{et.\ al.} \cite{combinedhuang01,combinedhuang2} also used variable kinetics, but studied equilbrium linear polymers with $c_{st} = 1$ rather than network-forming APs with $c_{st} \ll 1$.
Details of the Huang \textit{et.\ al.} method are discussed extensively in Ref.\ \cite{combinedhuang01}.
The key differences of our method from Ref.\ \cite{baljon07} are the imposition of binary bonding and the use of variable $\tau_{MC}$ (they used only one $\tau_{MC} = 0.2\tau_{LJ}$).
Another difference was that Refs.\ \cite{baljon07,combinedhuang01,combinedhuang2} all used a much stronger thermostat, giving overdamped (Brownian) dynamics.

Many previous studies have used nonspecific (e.\ g.\ strengthened attractive Lennard-Jones or Coulombic) interactions which allow SMs to form arbitarily many simultaneous SBs \cite{nguyen95,khalatur99,kumar99,kumar01,ayyagari04,guo05,combinedloverso,manassero05,goswami07}.
This results in formation of interesting structures such as micelles and micelle-bridge networks, which occur in real AP systems such as associating ionomers (see e.\ g.\ Ref.\ \cite{combinedpitsikalis}).
In contrast, the (experimental) APs we wish to model tend to form networks more like classical rubbers.

Most previous studies \cite{groot94,liu96,khalatur99,kumar01,loverde05,baljon07} varied temperature $T$ at fixed SM bonding strength.
This does not isolate the effects of $T$ on associative bonding from its other effects such as the dynamical slowdown which occurs in normal (non-associative) polymers. 
To get a full picture of AP physics, one should vary both $h$ and $T$ independently \cite{combinedhuang01}.
We follow this approach.

Other differences from previous simulation studies are more associated with the systems employed than the methods applied.
Many studies have considered only telechelic chains \cite{khalatur99,combinedloverso,loverde05,guo05,ayyagari04,manassero05,baljon07,goswami07}.
Telechelics are appealing in their simplicity, but their network-forming abilities are naturally limited;
for binary bonding, at least 3 SMs/chain are required to form good networks.
Weakly entangled chains ($N \sim N_e$) may be ideal \cite{deGreef08} for technological goals such as enhanced melt processability at high $T$ and network strength at low $T$.
The majority of previous studies have employed extremely short $N \ll N_e$ chains \cite{manassero05,liu96,khalatur99,ayyagari04,guo05,baljon07}, but we consider systems with $N \sim N_e$.
Finally, the majority of previous studies have focused on small $\rho$ corresponding to solutions \cite{nguyen95,khalatur99,kumar99,kumar01,guo05,baljon07,combinedhuang01,combinedhuang2}.
AP solutions exhibit a wide range of intriguing phenomena, in particular competition between gelation and phase separation \cite{groot94,liu96}, which, however, we do not wish to consider here.
In addition, the presence of solvent can dramatically weaken the effective strength of sticky bonds in real systems \cite{sontjens00,yamauchi03,deGreef08}; this effect is beyond the scope of our model.
We therefore focus on systems with $\rho$ corresponding to a dense pure melt with no solvent.\newline

Mappings of the bead-spring model to real, dense polymer melts \cite{kremer90} produce different $\tau_{LJ}$ in the range $10^{-10.5\pm1.5}s$. 
Present day computers can achieve runs (for the system sizes used here) of up to $\sim 10^{7}\tau_{LJ} \sim 10^{-3.5\pm1.5}s$, but runs this long can not be performed over a broad parameter space.
In contrast, sticky bond lifetimes in experimental systems are typically at least $10^{-4}s$, and often many orders of magnitude longer \cite{sijbesma97,sontjens00,yount05}.
Thus any attempt to capture specific SM chemistries and at the same time use systems large enough to study bulk dynamics and mechanical properties would exceed the capabilities of present day supercomputers \cite{foot8}.
Coarse-grained modelling with the goal of studying the dynamics of AP systems \textit{by analogy} is the only currently feasible approach for bulk systems, so we make no attempt to mimic specific chemistries.
The only published simulations of which we are aware that model AP networks with specific chemistries \cite{wang08,genesky08} are pure Monte Carlo studies that used a very coarse-grained (lattice) bond-fluctuation model \cite{karmesin88} and focused on static properties.

\subsection{Static Properties: Validation of Hybrid MD/MC Method}
\label{subsec:twostate}

As discussed above, systems contain a total of $N_{st} = N_{ch}N c_{st}$ sticky monomers.
Due to the binary bonding rules, the maximum number of sticky bonds that can exist in the system at any given time is $N_{st}/2$.
If the probability that an SM is bound into an SB is $p_{active}$, then the total number of SBs in the system is $N_{st}p_{active}/2$.
If $A$ represents an unbound SM and $A_2$ represents a bound SM pair, these factors define the concentrations 
\begin{equation}
\begin{array}{rcl}
\left[A\right] & \equiv & \rho c_{st} (1 - p_{active}),\\
&  & \\
\left[A_{2}\right] & \equiv & \rho c_{st} p_{active}/2, 
\end{array}
\label{eq:concdefs}
\end{equation}
where square brackets denote concentrations.
If the equilibrium value of $p_{active}$ is $p^{*}$, then the equilibrium constant for SB association is defined (by the law of mass action for the reaction $A + A \leftrightarrow A_2$) as
\begin{equation}
\begin{array}{rcccl}
K_{eq} & \equiv & \displaystyle\frac{[A_2]}{[A]^2} & \equiv & \displaystyle\frac{p^{*}}{2\rho c_{st} (1-p^{*})^2}
\end{array}
\label{eq:Keqdefn}
\end{equation}
for binary bonding.

\begin{figure}[h]
\includegraphics[width=3.375in]{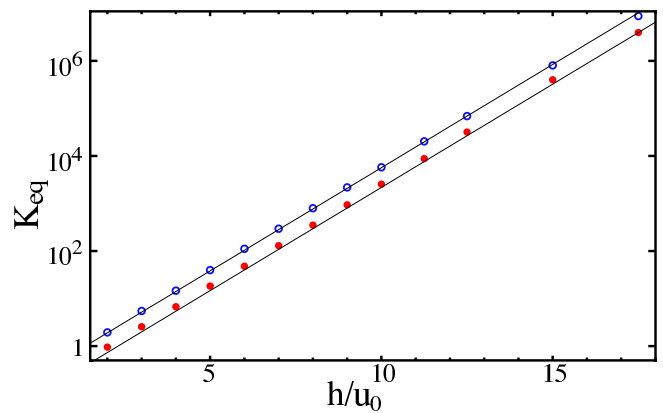}
\caption{Sticky association in equilibrium; simulation data and test of Equation \ref{eq:alphabetah}. All results are for $N_{ch}N = 280000$, $c_{st} = 0.08$ systems with $k_B T = 1.0u_0$ and $\tau_{MC} = 1.0\tau_{LJ}$.  Closed circles are simulation values of $K_{eq}$ from Eq.\ \ref{eq:Keqdefn} for $N=50$ polymers and open circles are for $N=1$ dimer-forming systems.  The straight lines are exponential fits, to Eq. \ref{eq:alphabetah}, for $K_{eq}^{TS}$.}
\label{fig:valid1}
\end{figure}

Figure \ref{fig:valid1} shows simulation data in which $p^{*}$ was evaluated from equilibrated simulations at fixed $h$ and $K_{eq}$ obtained from Eq.\ \ref{eq:Keqdefn}.   
Circles show values of $K_{eq}$ for $N=1$ and $N=50$ systems.
As expected, $K_{eq} \sim exp(h/k_B T)$.
The data shown are for $\tau_{MC} = 1.0\tau_{LJ}$, but we have verified that $p^{*}$ is independent of $\tau_{MC}$ (to within statistical errors) for all $h$ tested, over the range $\tau_{LJ} \leq \tau_{MC} \leq 100\tau_{LJ}$.
Because there is an entropy cost $\sim k_B T$ to form a SB, few SBs form for $h < 2u_0$.
As $h/u_0$ ranges from $2$ to $17.5$, the equilibrium constant $K_{eq}$ varies over more than six orders of magnitude, from $0.96$ to $3.9\cdot10^6$. 
This is a wider range of $h$ and $K_{eq}$ than considered in previous simulation studies.
A standard \cite{pandey85} finite size scaling analysis of the percolation gel transition is given in Appendix \ref{app:percgt}. 
For $k_B T = 1.0u_0$, percolation occurs at $h = h_{perc} = 4.25u_0$, so we consider values of $h$ up to $\sim 4$ times above the gelation transition.

Note that the $\tau_{MC}$-independence of $p^{*}$ allows systems to be equilibrated efficiently using a low $\tau_{MC} = \tau_{LJ}$.
Higher values of $h$ (for polymeric systems) are impossible to equilibrate on present-day computers with our current method; equilibration is discussed further in Section \ref{sec:results}.
However, the highest values of $K_{eq}$ considered here are comparable to those observed in some experiments on multiple-H-bonding SMs \cite{sontjens00,rotello08}.

Data from multiple system sizes are also useful in further validating the simulation model.
Ben-Naim and Krapivsky have pointed out that systems which reversibly polymerize undergo a nonthermodynamic gelation transition \cite{bennaim08} when the fragmentation (in our case, SB breaking) process is too weak.
The average number of clusters (aggegrates) at any given time is $N_{agg} \equiv N_{ch}/N_n$, where $N_n$ is the number-averaged cluster size (Appendix \ref{app:percgt}).
$N_{agg} \equiv N_{ch}$ in the absence of sticky bonding and $N_{agg}\to 1$ in the limit of large $h$, because all the chains combine into a single network (as in an ideal rubber).
Our systems, in the terms of Ref.\ \cite{bennaim08}, are ``thermodynamic'' if and only if: (1) $N_{agg}$ is linearly proportional to $N_{ch}$ below percolation (i. e. for $h < h_{perc}$) and (2) the probability distibution of cluster sizes $P(M)$ (Appendix \ref{app:percgt}) is independent of $\tau_{MC}$.
An arbitrary simulation method will not necessarily display a `thermodynamic' gel transition; failure to do this would be a serious flaw according to our goals.
We therefore have verified that our model satisfies conditions (1) and (2) for $\tau_{LJ} \leq \tau_{MC} \leq 100 \tau_{LJ}$, and therefore properly captures reversible gelation.
Satisfaction of these conditions appears equivalent to the above-verified condition that $p^{*}$ is independent of $\tau_{MC}$ \cite{foot16}.

\begin{figure}[h]
\includegraphics[width=3.375in]{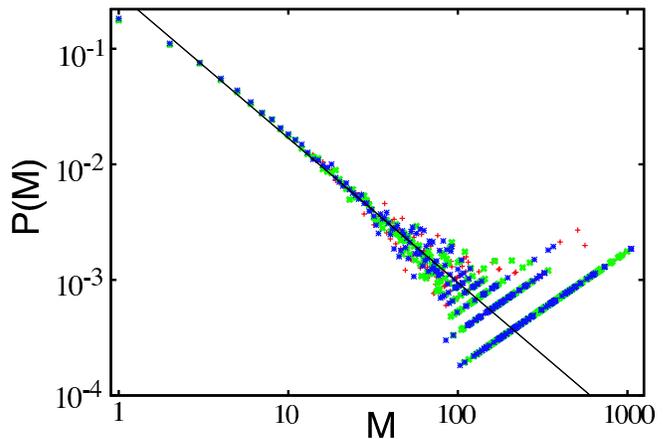}
\caption{Cluster size distribution.  $P(M)$ is the probability that a chain will be part of a disconnected cluster of $M$ chains (i. e. the weight fraction of $M$-clusters).  All results are for $N = 50$, uniform $c_{st} = 0.08$ systems with $k_BT = u_0$ and $h = 4u_0$.  Data for different kinetic rates are shown: $\tau_{MC}/\tau_{LJ} = 1$ (blue stars), 10 (green $\times$), and 100 (red +).  The upward slope at large MW is due to the statistics of small numbers.  Results are averaged over 100 statistically independent samples.}
\label{fig:MWdistrib}
\end{figure}

Figure \ref{fig:MWdistrib} shows data for $P(M)$ at $h=4u_0$ and $k_B T=1.0u_0$ (i.\ e.\ just below percolation).
The collapse of the data shows \cite{bennaim08} that cluster formation/dissociation is an equilibrium processes and supports our arguments that the algorithm satisfies detailed balance for the range of $\tau_{MC}$ considered here.
Also, $P(M)$ shows some interesting properties which demonstrate that our modelled systems form good (rubber-like) networks.
The line shows a fit to a $P \propto (M)^{-5/4}$ power law, which is consistent with the fractal dimension $D_{frac} = 4$ of aggregrates and the expected power law $ln(P) \sim -(1 + 1/D_{frac})ln(M)$ for networks \cite{milchev00}.
In contrast, dense telechelic systems have an exponential $P(M) \sim exp(-M/<M>)$ distribution. 
The absence of any large exponential contribution in our $P(M)$ at large $M$ indicates that long linear clusters are not common.
Therefore, though our parent chains only contain 4 SMs each, we are confident that that is enough to accurately capture AP network physics.

\subsection{SB Dynamics and Two-State Model}
\label{subsec:twostatedynamics}

\begin{figure}[h]
\includegraphics[width=3.375in]{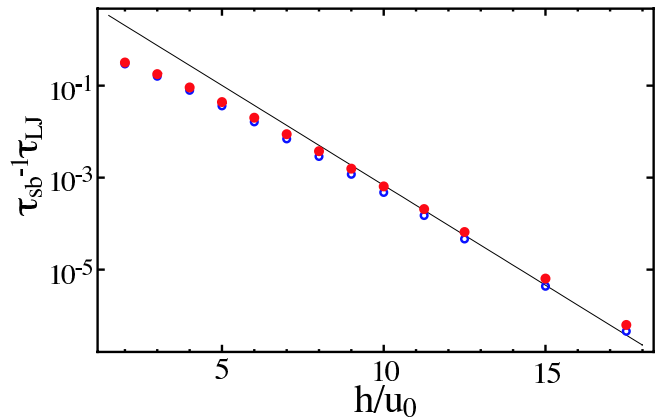}
\caption{Sticky bond lifetimes. All results are for 280000-bead, $c_{st} = 0.08$ systems with $\tau_{MC}=1.0\tau_{LJ}$ and $k_B T=u_0$.  Closed circles are simulation data for $N=50$ polymers, open circles are data for $N=1$ dimer-forming systems, and the straight line is an exponential `fit', shown only as a guide to the eye.}
\label{fig:baretausb}
\end{figure}

Figure \ref{fig:baretausb} shows simulation results for the average sticky bond lifetime, $\tau_{sb}$, in quiescent systems at chemical equilibrium.
Simple thermal activation of SB dissociation would suggest exponential behavior, $\tau_{sb}^{-1} \propto exp(-h/k_B T)$.
In fact the results are markedly nonexponential.
Interestingly, SB lifetimes in polymeric systems are (apparently) always lower than those in dimer-forming systems.
This is consistent with differences in chain connectivity; SBs embedded in polymers experience additional `pulling' forces due to transmission of the random thermal forces (which produce diffusive motion) through covalent bonds along their parent chains.
Additional reductions in $\tau_{sb}$ could potentially arise from increased steric hindrance to bonding for embedded SMs.
While this ``polymeric'' effect on $\tau_{sb}$ should dependent sensitively on $N$, $c_{st}$ and $T$, to our knowledge it is not included in any theories for AP networks.

\begin{figure}[htbp]
\includegraphics[width=3.25in]{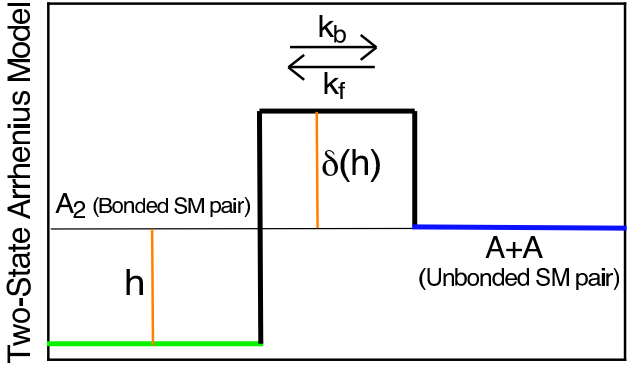}
\caption{Arrhenius 2 state model.  Refs.\ \cite{combinedhuang01,combinedhuang2} did not introduce an $h$-dependent $\delta$.}
\label{fig:2state}
\end{figure}

The simulation data in Figures \ref{fig:valid1} and \ref{fig:baretausb} can be better understood by mapping the Monte Carlo procedure and $U_{sb}(r,h)$ onto a two state Arrhenius model for sticky bonding.
The model is depicted in Figure \ref{fig:2state}.
Bonded SM pairs are assumed to have an energy $-h$, unbonded SMs have zero energy, and we introduce an $h$-dependent barrier $\delta(h)$.

The Monte Carlo rules described in Section \ref{subsec:descrip} allow us to assume that sticky bond formation obeys second order kinetics and dissociation obeys first order chemical kinetics, as they should as long as $\rho c_{st} \ll 1/a^{3}$ \cite{atkins78,rubinstein98}.
The SB formation/dissociation process can be represented as the chemical reaction
\begin{equation}
\begin{array}{rcl}
A + A & \stackrel{\overset{k_f}\longrightarrow}{\underset{k_b}\longleftarrow} & A_2
\end{array}
\label{eq:formbreak}
\end{equation}
where $k_f$ and $k_b$ are the rate constants for SB formation and dissociation.
Then the equation for chemical equilibrium is 
\begin{equation}
\begin{array}{rcl}
k_f [A]^2 & = & k_b [A_2].
\end{array}
\label{eq:chembalance}
\end{equation}  
In the Arrhenius two state model the rate constants are given by:
\begin{equation}
\begin{array}{rcl}
k_f & = & \alpha \exp(-\delta(h)/k_B T),\\
& & \\
k_b & = & \beta \exp(-(h + \delta(h))/k_B T),\\
\end{array}
\label{eq:twostaterateconsts}
\end{equation}
where $\alpha$ and $\beta$ are constants with dimensions of volume$\times$frequency and frequency, respectively.
Note that the above is a ``mean field'' model \cite{cates87} in that it ignores correlations between sticky monomers (i.\ e.\ concentration fluctuations).
Thus $k_f$ and $k_b$ (and especially $\alpha$ and $\beta$) will in general depend \cite{loverde05} on $N$, $\rho$, $c_{st}$, and (through second order effects such as the variation of $\rho$ at fixed pressure) $T$.  

In thermal equilbrium, Eq.\ \ref{eq:chembalance} gives the equilibrium constant
\begin{equation}
K_{eq}^{TS} \equiv \displaystyle\frac{k_f}{k_b} = \displaystyle\frac{\alpha}{\beta}exp(h/k_B T).\label{eq:alphabetah}
\end{equation}
Eq.\ \ref{eq:alphabetah} fits simulation results for $K_{eq}$ very well, as shown in Fig.\ \ref{fig:valid1}.
In AP networks at even higher values of $h$, Eq.\ \ref{eq:alphabetah} should fail due to `trapped' open SMs \cite{loverde05} that cannot find partners, but this effect is negligible for the systems considered here.

We now compare two state model predictions to simulation data and map the latter to the former.
In the two state model, the mean SB lifetime $\tau_{sb}$ is just $\tau_{sb} = k_b^{-1}$.
Similarly, the probability that an unbonded pair in the `2A' state (Fig.\ \ref{fig:2state}) will jump over the barrier is just $exp(-\delta(h)/k_B T)$. 
This is also the success rate $S_{MC} \equiv k_f/\alpha$ for Monte Carlo SB formation attempts, so $\delta(h)$ can be directly measured from the simulations: $\delta(h)/k_B T = -ln(S_{MC})$.

\begin{figure}[h]
\includegraphics[width=3.275in]{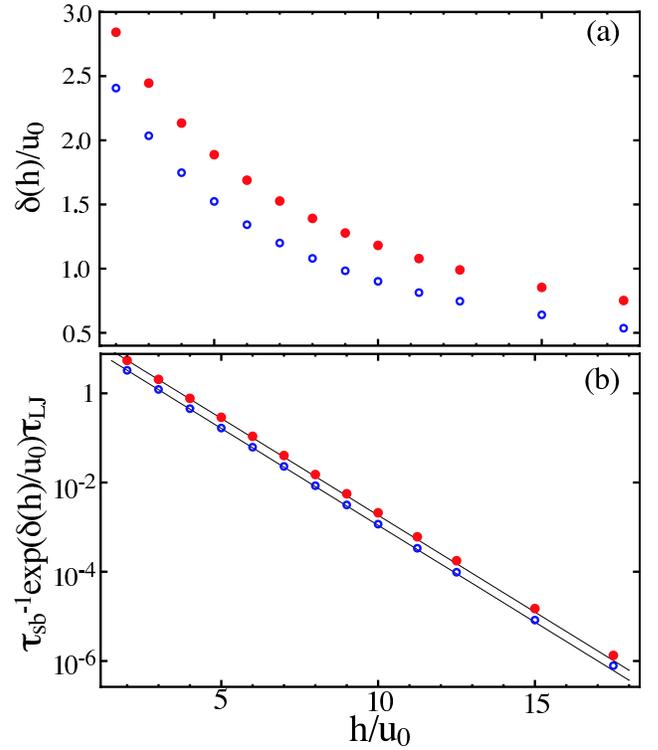}
\caption{Validation of method. All results are for 280000-bead, $c_{st} = 0.08$ systems with $\tau_{MC}=1.0\tau_{LJ}$ and $k_B T = u_0$.  Closed circles are simulation data for $N=50$ polymers, open circles are data for $N=1$ dimer-forming systems, and  straight lines are exponential fits.   Data shown are: (a) $\delta(h)/k_B T$, and (b) $\tau_{sb}^{-1}\exp(\delta(h)/k_B T)$.}
\label{fig:valid2}
\end{figure}

In Figure \ref{fig:valid2}, panel (a) shows simulation results for $\delta(h)$ and panel (b) shows simulation results for  $\tau_{sb}^{-1}\exp(\delta(h)/k_B T)$.  
The latter shows that the perfect exponential decay expected from Eq.\ \ref{eq:twostaterateconsts}, $\tau_{sb}^{-1}\exp{(\delta(h)/k_B T)} = k_b = \beta \exp{(-h/k_B T)}$, is actually observed.
Note that this Arrhenius behavior \textit{was in no way imposed}; it emerges naturally, showing the utility of the two state model in understanding the behavior of our simulations.

The parameters $\alpha$ and $\beta$ can be extracted from the data in Figs.\ \ref{fig:valid1} and \ref{fig:valid2}. 
For $\tau_{MC} = \tau_{LJ}$, $\alpha = 6.3a^3/\tau_{LJ}$ and $\beta = 24/\tau_{LJ}$ for dimers, while $\alpha = 4.0a^{3}/\tau_{LJ}$ and $\beta = 41/\tau_{LJ}$ for $N=50$ chains.
The smaller $\alpha$ measured for polymer-embedded SMs is consistent with the above-hypothesized increased steric constraints.
The large rate constant $\beta = 41/\tau_{LJ}$ indicates a potential problem with the simulations.
$\beta$ is an effective ``attempt frequency'' for breaking sticky bonds, which implies that the MC timestep $\tau_0$ should be small compared to $\beta^{-1}$.
Larger $\tau_0$ will in principle produce systematic errors.
The data shown above are for $\tau_0 = 1.0\tau_{LJ}$, which is large compared to $\beta^{-1}$.
Simply reducing $\tau_0$ is problematic because it sharply reduces the parallel efficiency of the simulations.

However, we have used values of $\tau_0$ as small as $.05\tau_{LJ}$, and find that all errors produced by using $\tau_0 = 1.0\tau_{LJ}$ are small in quiescent systems at equilibrium; for example, the systematic error in $\tau_{sb}$ at $h=10u_0$ is about 1\%.
While the errors in dynamical properties are somewhat larger at small $h$ ($h \ll 10u_0$), in this paper we focus on dynamics for $h \geq 10u_0$ and use $\tau_0 = 1.0\tau_{LJ}$.
In all cases, all differences produced by smaller $\tau_0$ are small compared to the differences between systems contrasted in Section \ref{sec:results}, and comparable to our statistical errors, i.\ e.\ $\sim 1\%$.
For nonequilibrium systems, however, systematic errors are larger.
Thus all nonequilibrium and mechanical-property tests in this paper are performed using $\tau_0 \leq 0.2\tau_{LJ}$.

In summary, to within our noise, increasing $\tau_{MC}$ leaves the static properties of our model AP networks (Figure \ref{fig:valid1}) unchanged, and changes the sticky bonding dynamics (Figure \ref{fig:valid2}a-b) only through the prefactor $\tau_{sb} \propto \tau_{MC}^{-1}$.
The role of $\tau_{MC}$ in the dynamics therefore appears in the rate constants $\alpha$ and $\beta$, which are also proportional to $\tau_{MC}^{-1}$.
Increasing $\tau_{MC}$ slows down the chemical kinetics of the SBs (both formation and dissociation) relative to the underlying polymeric time scales, while leaving the thermodynamics unchanged.
This is why we claim our model can separate thermodynamics and kinetics.
The variation of $h$ and $\tau_{MC}$ employed here may be thought of as corresponding to ``scanning'' across chemically different sticky monomers.
Given the time scale problem mentioned above, this scanning is only qualitative.  
However, we show below that it is very useful in understanding AP systems.

\section{Results}
\label{sec:results}

Previous work has shown  \cite{kumar01,baljon07} that the most dramatic changes in dynamics, our primary interest, take place not at $h_{perc}$ but rather at considerably higher $h$.
The rest of this paper considers systems with $h \gg h_{perc}$ and $p^{*} \gtrsim 0.95$.
This is the ``physical gel'' regime \cite{kumar99} where nearly all chains are (at any moment) part of a single aggregate.
A ``snapshot'' of a physical gel looks much like a crosslinked rubber, yet chains are delocalized and the system can flow at long times.
One of the most interesting properties of physical gels is their transition to chemical gels as $h$ increases or $T$ decreases.
In this ``physical-chemical gel transition'' (PCGT), chains become localized \cite{kumar01,baljon07} in a manner analogous to the ``caging'' effect produced upon cooling fragile glass-forming systems \cite{donati99,foot15}.

For the $N$, $T$, and $c_{st}$ considered here, the PCGT occurs \cite{foot6} at bonding strength $h_{PCGT} > 17.5 u_0$.
The broad range ($5u_0 \lesssim h \lesssim 17u_0$) between the percolation (Appendix \ref{app:percgt}) and localization transitions is consistent with the findings of Kumar and Douglas \cite{kumar01} as well as Baljon \textit{et.\ al.} \cite{baljon07}, who both, however, used constant SB strength and varied $T$.
The broad range is not dependent on having only a few sticky monomers per chain, although increasing $Nc_{st}$ at fixed $\rho$ will broaden the range by lowering $h_{perc}$.
Here we focus on values of $h$ which are well below $h_{PCGT}$, and thus ``in the middle'' of the physical gel regime.

Another of the key features of physical AP gels is sticky bond recombination.
The concentration $\rho c_{st}(1 - p^{*})$ of free SMs is small.
Moreover, the motion of free SMs is constrained by their (transiently but usually) bonded intrachain neighbors.  
Thus SM pairs tend to recombine after SB-dissociation events.
This leads to to a second characteristic timescale for individual sticky bonds; in addition to the ``bare'' lifetime $\tau_{sb}$, there is \cite{rubinstein98} a larger, ``effective'' SB lifetime $\tau^{*}$, which can be thought of as the average time for initially bonded SMs to 	``separate'' (i.\ e.\ no longer recombine) as opposed to merely debond.
It is of interest because rheological experiments typically measure $\tau^{*}$; $\tau_{sb}$ is more difficult to access \cite{cates87,rotello08,katieunpublished}.
Values for $\tau_{sb}$ and $\tau^{*}$ (defined more specifically in Appendix \ref{app:SBrecomb} and discussed further in Section \ref{subsec:crossover}) for a wide variety of systems are given in Table \ref{tab:htauMCtausb}.
We have already shown how $\tau_{sb}$ is affected by polymer physics - indirectly through covalent backbone bonds.
Now we study the ways in which SB recombination influences and is influenced by the interplay of SB thermodynamics, SB kinetics, and polymer physics.
We perform our study in terms of measurements of diffusion, $\tau^{*}$, dynamical heterogeneity, nonequilibrium chemical dynamics, and nonlinear mechanical properties.
All results presented below are for systems that were first equilibrated for many $\tau^{*}$.
As will be shown, these are best understood by determining whether SB recombination is diffusion-limited or kinetically limited.

\subsection{Diffusion}
\label{subsec:diffusion}

\begin{figure}[htbp!]
\includegraphics[width=3.3in]{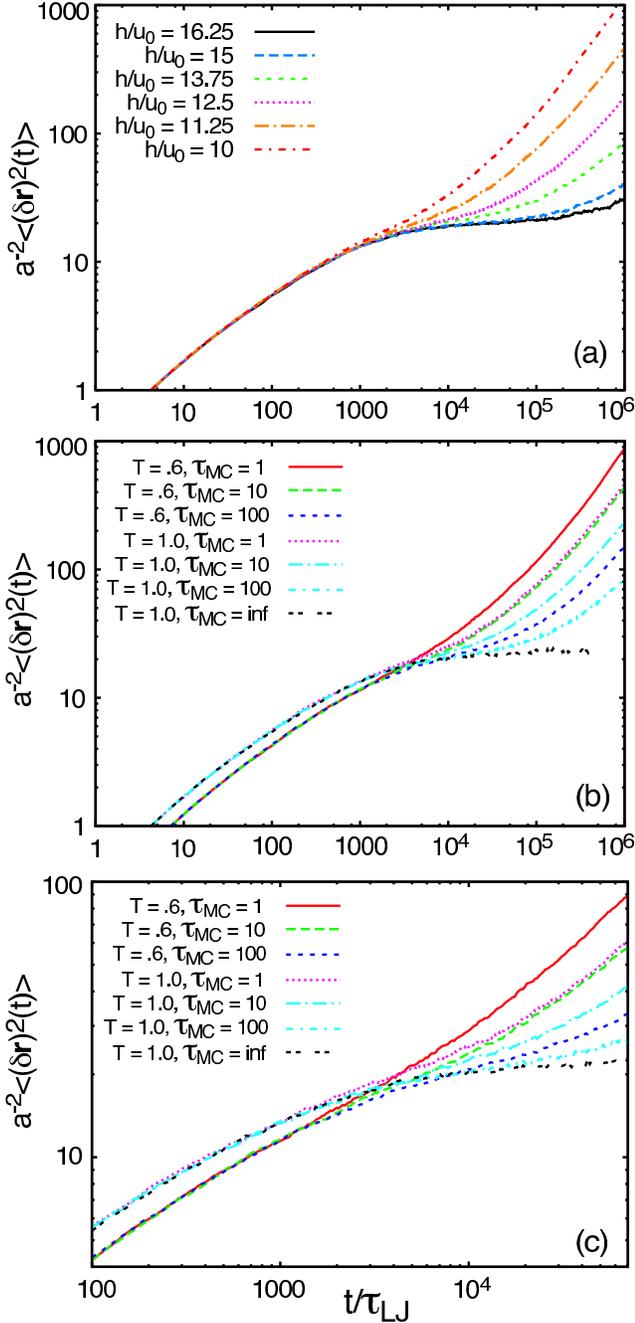}
\caption{Diffusion as function of $h$, $T$, and $\tau_{MC}$.   All systems have $N_{ch}=1400$.  Panel (a): $k_B T = u_0$, increasing $h$. Panel (b):  $h/k_BT =11.25$; $k_B T=u_0$ and $k_B T=0.6u_0$, increasing $\tau_{MC}$. Panel (c) is a blowup of (b) showing the crossover.   Lines from top to bottom for each $T$ in panels (b-c) are for $\tau_{MC}/\tau_{LJ} = 1$, $10$, $100$, and (for $k_B T=1.0u_0$) $\infty$.   The legends for panels (b-c) give $T$ in units of $u_0/k_B$ and $\tau_{MC}$ in units of $\tau_{LJ}$.}
\label{fig:h1125diff}
\end{figure}

The effect of varying different thermodynamic and kinetic parameters on monomer diffusion (mean squared displacement $<(\delta \vec{r})^{2}(t)>$) is shown in Figure \ref{fig:h1125diff}.
Panel (a) shows the variation as $h$ is increased at $\tau_{MC}=1.0\tau_{LJ}$ and $k_B T=1.0u_0$.
At short times ($t \ll \tau_{sb}$), results for different values of $h$ collapse, showing (as expected) that sticky bonding has little effect on diffusion on these time scales.
At larger times ($t \gtrsim \tau_{sb}$) results show a progressive localization and `caging' effect, similar to that described in Refs.\ \cite{kumar01,baljon07}, as sticky bond strength increases. 
At $h = 10u_0$, little localization occurs because \cite{rubinstein98} $\tau_{sb}$ is less than the Rouse time of the chains in the absence of sticky bonding ($\tau_{R} \simeq 2.6\cdot10^{3}\tau_{LJ}$). 
As $\tau_{sb}$ increases with increasing $h$, the curves develop a ``shoulder'' which illustrate the temporary caging associated with physical gels.
This temporary cage becomes permanent as $\tau_{sb}\to \infty$ (as in a classical crosslinked rubber).
Data in panel (a) support our earlier statement that this occurs for some $h \gtrsim 17u_0$.

Panels (b-c) show a pair of interesting effects.  
First, for $h = 11.25u_0$ and $k_B T=1.0u_0$, increasing $\tau_{MC}$ has the same qualitative effect as increasing $h$ at fixed $\tau_{MC}$.  
Data for $<(\delta \vec{r})^{2}(t)>$ collapse for $t$ less than the smallest $\tau_{sb}$ (i.\ e.\ $\tau_{sb}$ for the lowest $\tau_{MC}$).
For longer times, the data develops a shoulder which increases in width as $\tau_{MC}$ increases.
Data for an equilibrated system with MC deactivated (i.\ e.\ $\tau_{MC} = \infty$) shows ``chemical gel'' (ideal rubber) behavior; chains are permanently localized.

Second, data from systems with the same ``SB thermodynamics'' (i.\ e.\ the ratio $h/k_BT = 11.25$) but different ambient conditions ($h = 6.75u_0$ and $k_B T = 0.6u_0$) shows interesting contrasts which illustrate the interplay of SB dissociation and underlying polymer physics.
For $t \lesssim \tau_{sb}$, data for $<(\delta \vec{r})^{2}(t)>$ still collapse, but data from the lower-$T$ systems collapse on a lower value.
This is not at all surprising, as polymeric diffusion is well known to slow with decreasing $T$.
However, though $h/k_B T$ is the same, values of $\tau_{sb}$ are smaller for the lower-$T$ systems, perhaps because $h$ is smaller and SB breaking is favorable at smaller $r$ (see Fig.\ \ref{fig:smpotential}) \cite{foot10}.
Thus the diffusion data actually cross over at intermediate time scales (panel c) and the lower $T$ systems show greater mobility at fixed $h/k_B T$, a most unusual state of affairs.
While the case presented here is somewhat artificial because in a real polymer melt $\rho$ would decrease with $T$ and lead to further diffusive slowdown, we believe the point that varying $T$ at fixed SB thermodynamics should change relaxation at different timescales differently should be generally valid.
For example, the frequency ($\omega$) dependence of the dynamical moduli $G(\omega; T)$ \cite{ferry80} should change with $T$ in nontrivial ways.
In other words, time-temperature superposition should be violated.

It is useful to relate the mean squared displacement to the cage size $a_{cage}$ and ``escape parameter'' $f(t)$ using the definition
\begin{equation}
<(\delta \vec{r})^{2}(t)> \equiv a_{cage}^{2} f(t).
\label{eq:cagesubdiff}
\end{equation}
In the $\tau_{sb}\to\infty$ limit, $a_{cage}^{3} \sim (\rho c_{st})^{-1}$ is the volume explored by sticky  monomers \cite{foot17}.
The ``chemical gel'' time $t_{chem}$ is the time at which $f(t)$ approaches unity in this limit.
Data for the $\tau_{MC} = \infty$ system in Figure \ref{fig:h1125diff}(b)  (with  $h = 11.25u_0$, $N = 50$, and $c_{st}=.08$) shows that $a_{cage}^{2} \simeq 23a^{2}$ and $t_{chem} \sim 10^{4.5}\tau_{LJ}$.
For finite $\tau_{sb}$, one can define $t_{cage}$ as a ``caging'' time describing the (de)localization of SMs \cite{kumar01,baljon07}; $f(t)$ then has the general form $f(0) = 0$, $f(t) \sim 1$ for $t \sim t_{cage}$, and $f(t) > 1$ for $t > t_{cage}$.

\begin{figure}[h]
\includegraphics[width=3.375in]{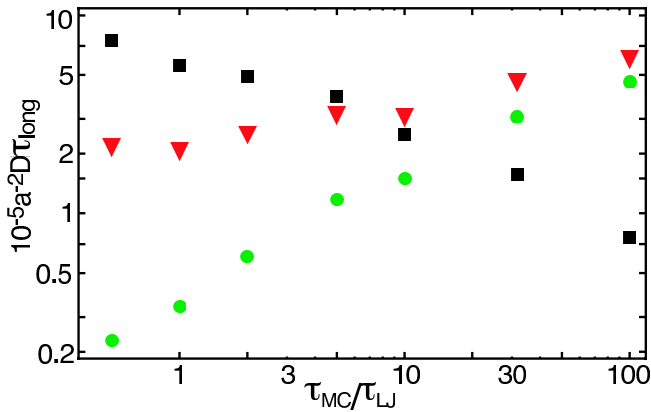}
\caption{Scaling of diffusion with various candidate ``long'' relaxation times.  Data shows $D\tau_{long}$ for $\tau_{long} = \tau_{sb}$ (circles),  $\tau_{long} = \tau^{*}$ (triangles), and $\tau_{long}=8\cdot10^{4}\tau_{LJ}$ (squares).  The last value is chosen so the ``bare'' diffusion constant $D$ can be shown on the same plot.  All results are for systems with $N_{ch}=1400$, $h=11.25u_0$ and $k_BT = 1.0u_0$. $D\tau_{MC}$ is not shown because $\tau_{sb} \propto \tau_{MC}$.}
\label{fig:dvstaustar}
\end{figure}

The monomeric diffusion constant $D$, as measured by $lim_{t\to\infty} <(\delta \vec{r})^{2}(t)> \sim 6Dt$, should vary inversely with some  ``long'' characteristic time $\tau_{long}$ of the system, roughly defined as the time for chains to diffuse by their end-end distance.
Candidates for $\tau_{long}$ include $\tau_{sb}$ and $\tau^{*}$.
Ref.\ \cite{rubinstein98} predicts $\tau_{long} \propto \tau_{sb}$ for weakly binding physical gels and $\tau_{long} \propto \tau^{*}$ in the strong-binding (near-chemical) limit.
Figure \ref{fig:dvstaustar} shows results for $D\tau_{sb}$, $D\tau^{star}$, and $8\cdot10^{4}\tau_{LJ}D$ from Table \ref{tab:htauMCtausb} for $h=11.25u_0$, $k_B T = 1.0u_0$ systems, over a wide range of $\tau_{MC}$.
$D$ decreases with increasing $\tau_{MC}$ slower than both $(\tau^{*})^{-1}$ and $\tau_{sb}^{-1}$, but it tracks the former more closely than the latter.
Thus results for this value of $h$ are apparently intermediate between the ``weak'' and ``strong'' physical gel limits described in Ref.\ \cite{rubinstein98}.

\subsection{Crossover between Diffusion-Limited and Kinetially Limited SB recombination}
\label{subsec:crossover}

Table \ref{tab:htauMCtausb} shows $\tau^{*}$ and $\tau^{*}/\tau_{sb}$ for all investigated systems.
As expected, increasing $h$ at fixed $\tau_{MC}$ increases $\tau^{*}/\tau_{sb}$, because for fixed kinetics recombination is more likely for thermodynamically stronger SBs. 
There are several possible regimes of possible relations between $\tau^{*}$ and $\tau_{sb}$ that can be related to diffusion (specifically, $<(\delta\vec{r})^{2}(t)>$) on intermediate timescales.
Systems with $\tau_{sb} \gg min(t_{cage},t_{chem})$ will exhibit kinetically limited sticky bond recombination (KL); in this regime $\tau^{*}/\tau_{sb}$ is predicted to be constant \cite{rubinstein98}.
However, if $\tau_{sb} \ll \tau_{cage}$, recombination will be dominated by ``correlated'' recombinations of SM pairs that have recently dissociated and have not had time to fully diffuse away from one another.
This is diffusion limited sticky bond recombination (DL).
To our knowledge, the DL regime and especially the crossover between DL and KL have not been previously studied for AP networks.

\begin{figure}[h]
\includegraphics[width=3.375in]{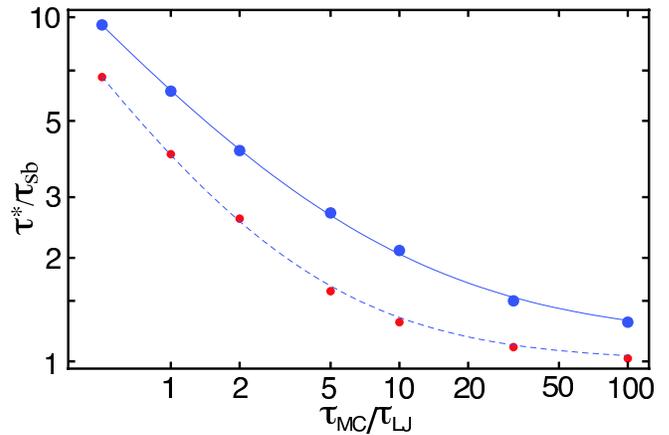}
\caption{Crossover from diffusion limited to kinetically limited sticky bond recombination for $h = 11.25$. All systems have $70000$ beads, with $c_{st}=.08$.  The upper data set is for $N=50$ polymers and the lower is for $N=1$ dimer-forming systems.   The fastest-kinetic systems ($\tau_{MC} = 0.5\tau_{LJ}$) use $\tau_0 = 0.5\tau_{LJ}$, and all others use $\tau_0 = \tau_{LJ}$.  The solid line shows a fit to Eq.\ \ref{eq:kltlpowerlaw} with $C = 1.15$, $K = 4.98\tau_{LJ}^{x}$ and $x = 0.74$.  The dashed line shows a fit with $C$ set to 1, $K=2.97\tau_{LJ}^{x}$ and $x=.94$.}
\label{fig:kltlcrossover}
\end{figure}

Figure \ref{fig:kltlcrossover} shows the variation of $\tau^{*}/\tau_{sb}$ with chemical kinetics for $h = 11.25u_0$ systems at $k_B T = 1.0u_0$.
Over a range of two orders of magnitude in $\tau_{MC}$, the data are well fit by the equation
\begin{equation}
\displaystyle\frac{\tau^{*}}{\tau_{sb}} = C + \displaystyle\frac{K}{(\tau_{MC})^{x}},
\label{eq:kltlpowerlaw}  
\end{equation}
where $C$ is the probability of recombination in the KL limit and $K$ is the contribution from diffusion-limited recombination.  
The exact form of Eq.\ \ref{eq:kltlpowerlaw} is not of great consequence; what matters is the broad crossover between regimes and the large change of $\tau^{*}/\tau_{sb}$ as a function of kinetics.
Nevertheless, since $\tau_{sb} = k_b^{-1}$, Eq.\ \ref{eq:kltlpowerlaw} can be interestingly rewritten
\begin{equation}
\tau^{*} = \displaystyle\frac{C + K\tau_{MC}^{-x}}{k_b}.
\label{eq:kltlrewrite}
\end{equation} 
The significance of Eq.\ \ref{eq:kltlrewrite} is its prediction of a nonlinear dependence of $k_{b}\tau^{*}$ on the rate constant $k_b$ for dissociation; $\tau_{MC}^{-x} \propto k_b^{x}$ (see Section \ref{subsec:twostatedynamics}).

$C$ and $K$ are of course not universal constants, but will depend on $h$, $\rho$, $c_{st}$, $T$, and $N$. 
In practice, one would expect $K\tau_{sb}^{x} \ll C$ when $\tau_{sb} \gg min(t_{cage},t_{chem})$.
More physically, the condition $K\tau_{sb}^{x} \ll C$ \textit{defines} the KL regime, where sticky bond reactions become ``mean-field'' in the sense of Cates \cite{cates87}.
The data in Figure \ref{fig:kltlcrossover} show that one can (at least in our model systems) move from the KL to the DL regimes simply by speeding up the chemical kinetics, if one is in the regime where $\tau_{sb}$ is comparable to the underlying polymer relaxation times such as $t_{chem}$.

For our systems, values of $C$ are close to values of $\tau^{*}/\tau_{sb}$ in our ``kinetically slow'' ($\tau_{MC}=100\tau_{LJ}$) systems, cf.\ Table \ref{tab:htauMCtausb}.  
$C$ increases with $h$ (qualitatively) as predicted by Rubinstein and Semenov \cite{rubinstein98}.
However, while our kinetically slow systems all have $C < 2$, the ``strong physical gel'' theory in Ref.\ \cite{rubinstein98} assumes $C \gg 1$, so we defer a detailed comparison to that theory to later work.
Here we merely make the positive observation that the basic prediction \cite{rubinstein98} of effective SB lifetime renormalization ($\tau_{sb} \to \tau^{*}$) works well at these relatively small $C$ and (somewhat surprisingly) over the entire studied KL$\to$DL crossover regime.
The renormalization $\tau_{sb} \to \tau^{*}$ accurately captures effective SB dissociation over a very broad parameter space \cite{foot11}.

On the other hand, an observation apparent from Fig.\ \ref{fig:kltlcrossover} is that SB recombination is in general only partly `polymeric' in nature.
In the limit of fast kinetics, values for $\tau^{*}/\tau$ in $N=1$ systems are nearly as high as for $N=50$ systems.  
However, the lack of a network prevents any caging effects, and so $\tau^{*}/\tau_{sb}$ decreases much faster with increasing $\tau_{MC}$, approaching unity (regardless of the value of $h$) at $\tau_{MC} = 100\tau_{LJ}$.
This illustrates that the crossover from DL to KL is analogous to a crossover between `dimeric' and `polymeric' recombination; in other words, chain connectivity (i. e. covalent bonding) becomes increasingly important as kinetics are slowed.
While the `dimeric' contribution to SB recombination cannot be simply ``subtracted out'' due to the different $\delta(h)$, these `dimeric-vs-polymeric' effects on SB recombination have been neglected by previous theories \cite{foot12}.

\subsection{Recombination and Dynamical Heterogeneity}
\label{subsec:dynheterog}

\begin{figure}[h]
\includegraphics[width=3.375in]{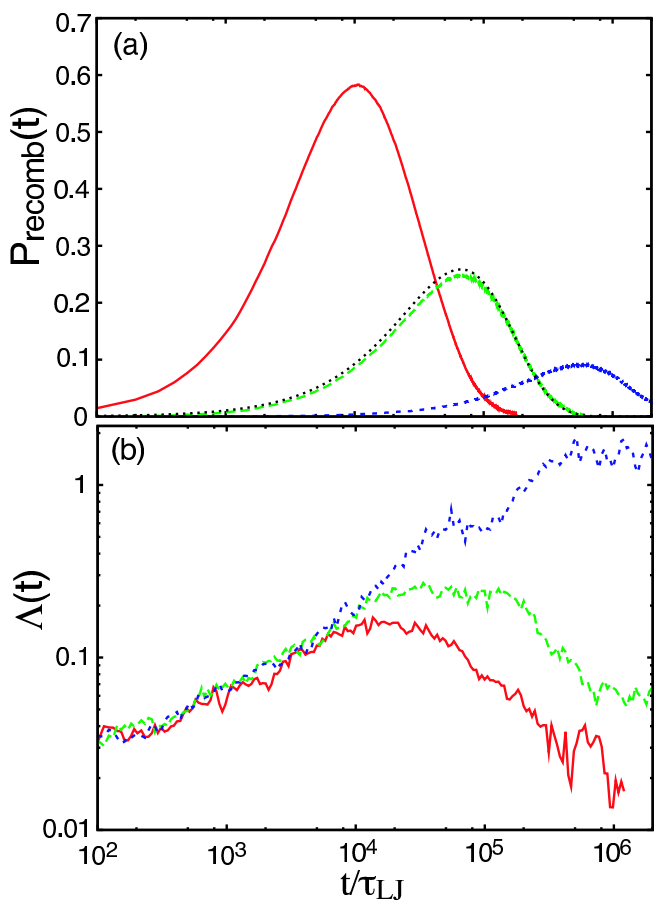}
\caption{SB recombination and dynamical heterogeneity.  All systems have $N_{ch}=1400$, $h=11.25u_0$ and $k_B T = 1.0u_0$.  Lines from top to bottom are for $\tau_{MC}/\tau_{LJ} = 1$, $10$, and $100$.  Panel (a): $P_{recomb}(t)$.  The black dotted line is $exp(-t/\tau^{*})-exp(-t/\tau_{sb})$ for $\tau_{MC}=10\tau_{LJ}$. Panel (b): $\Lambda(t)$.}
\label{fig:h1125deloc}
\end{figure}

Figure \ref{fig:h1125deloc}a shows simulation results for the SB recombination probability $P_{recomb}(t)$ (defined in Appendix \ref{app:SBrecomb}) for $h=11.25u_0$ systems with different chemical kinetics.
Although there is a small peak in $P_{recomb}$ (not displayed and low compared to the peaks shown in Fig.\ \ref{fig:h1125deloc}a) at very small times $t \sim \tau_{MC}$, the $t^{-5/4}$ behavior predicted \cite{oshaugh95} for the extreme diffusion-limited case is not found. 
This indicates none of our systems have ``too-fast'' kinetics \cite{combinedhuang01}.
For $t \gg \tau_0$, our results have the interesting form $P_{recomb}(t) \simeq \exp(-t/\tau^{*}) - \exp(-t/\tau_{sb})$; a comparison to actual data for $\tau_{MC} = 10\tau_{LJ}$ is shown. 

This form of $P_{recomb}(t)$ has a maximum at the ``delocalization'' time
\begin{equation}
\tau_{deloc} = \tau_{sb} \displaystyle\frac{y \log{y}}{y - 1},
\label{eq:taudeloc}
\end{equation}
where $y = \tau^{*}/\tau_{sb}$.
Bonded SM pairs trend towards moving away from each other after $\tau_{deloc}$.
Furthermore,
\begin{equation}
\displaystyle\frac{\tau_{deloc}}{\tau^{*}} = \displaystyle\frac{\log{y}}{y - 1},
\label{eq:newratio}
\end{equation}
which becomes small for $y \gg 1$.
This suggests $\tau_{deloc}$ (in addition to $\tau^{*}$) might be a key relaxation time in systems with large $y$.
However, this is speculative and needs further verification.

A useful measure of relaxation in complex fluids is the ``non-Gaussian'' parameter 
\begin{equation}
\Lambda(t) = \displaystyle\frac{3<\delta r^{4}(t)>}{5<\delta r^{2}(t)>^2} - 1,
\label{eq:nonGauParam}
\end{equation}
which is zero for normal diffusion and positive for systems where some particles move anomalously fast \cite{guo05}, particularly for ``hopping'' type motion. 
$\Lambda$ has been shown to be relevant to the structural relaxation of supercooled liquids and dynamical heterogenity \cite{donati99}.
The time $t_{deloc}$ at which $\Lambda$ is maximized and the maximum value $\Lambda_{max} = \Lambda(t_{deloc})$ both increase with decreasing $T$ in various systems, including associating polymers \cite{kumar01,guo05,baljon07}, as localization increases.
$t_{deloc}$ may be regarded as a crossover time after which the system begins to show liquidlike behavior.

Figure \ref{fig:h1125deloc}b shows the effect of kinetics on $\Lambda(t)$.
The effect of slowing kinetics at fixed $h/k_BT$ is similar to the effect of increasing $h/k_B T$ observed in previous studies \cite{kumar01,baljon07}.
It is interesting that increasing $\tau_{MC}$ increases dynamical heterogeneity.
The probable reason is that increasing $\tau_{MC}$, even though it leaves $p^{*}$ unaffected, decreases the likelihood of multiple closed SBs on the same chain breaking within a short time period. 
This is consistent with the idea \cite{leibler91} that coherent breaking of nearby SBs along a chain eases large-scale motion.
The increasing dynamical heterogeneity with increasing $t_{cage}$ is consistent with other results showing $\Lambda_{max}$ increases as localization ``transitions'' are approached, e.\ g.\ stretched-exponential relaxation of finite clusters \cite{ayyagari04}.

The data in Figs.\ \ref{fig:kltlcrossover}-\ref{fig:h1125deloc} also clearly show that $t_{deloc} \simeq \tau_{deloc}$ in systems where recombination is likely (i.\ e.\ when $\tau^{*}/\tau_{sb}$ is large compared to 1), and thus that delocalization is closely related to individual sticky bonds finding new partners in a ``hopping'' type motion.
However, these delocalization times are large compared to $t_{cage}$.
This is not surprising, as full delocalization should occur only when chains have lost all memory of their initial SB topology; this ``memory'' time is inherently polymeric in that it must increase with increasing $Nc_{st}$, similarly to a Rouse or reptation time \cite{doi86}.
Interestingly, the peaks of $\Lambda$ are broader than those of $P_{recomb}$.
This also likely arises either from cluster effects \cite{rubinstein98,ayyagari04} or other underlying many-SM phenomena that ultimately arise from the `polymer physics', i.\ e.\ the covalent connectivity of the parent chains.

\subsection{Nonequilibrium Chemical Dynamics}
\label{subsec:modelnoneq}

An important feature of our model is its ability to accurately capture the dynamics of systems in which  the sticky bonds are not in thermal equilibrium. 
The evolution of SB concentration is, following Equation \ref{eq:formbreak}, given by
\begin{equation}
\begin{array}{rcl}
\dot{[A_{2}]} & = & k_f [A]^{2} - k_b [A_2],
\end{array}
\label{eq:makebreakrates}
\end{equation}
which after plugging into Eq.\ \ref{eq:concdefs} and simplifying becomes
\begin{equation}
\dot{p}_{active} = 2k_f \rho c_{st} (1 - p_{active})^{2} - k_b p_{active}.
\label{eq:pactiveoft}
\end{equation}
Equation \ref{eq:pactiveoft} has an analytic solution.
For the special initial condition $p_{active}(t) = 0$ at $t = 0$, the solution is 
\begin{equation}
p_{active}(t) = d-\frac{\left(d^2-1\right) \tanh \left(2z\sqrt{d^2-1} t 
   \right) + d\sqrt{d^2-1}}{d \tanh \left(2z\sqrt{d^2-1} t \right)+\sqrt{d^2-1}},
\label{eq:specpoft}
\end{equation}   
where $z = \rho c_{st} k_f$ and $d = 1 + k_b/4z$.

\begin{figure}[h]
\includegraphics[width=3.375in]{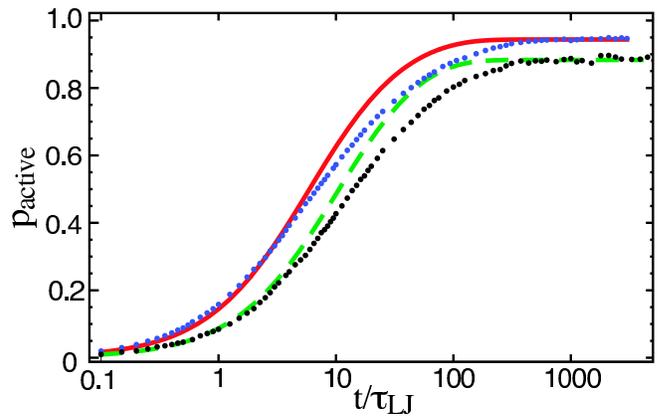}
\caption{Nonequilibrium capability of method: $h$-jump.  Solid (dashed) lines are the predictions of Eq.\ \ref{eq:specpoft} and upper (lower) circles show simulation data for $h=10u_0$, $k_BT = 1.0u_0$ ($h=6u_0$, $k_BT = 0.6u_0$) for an $N_{ch}=5600$ system after turning on sticky bonds.  Simulations used $\tau_0 = .05\tau_{LJ}$.}
\label{fig:pactivefig}
\end{figure}

Figure \ref{fig:pactivefig} compares this analytic prediction to simulation results for $p_{active}(t)$ upon activation of sticky bonding for two systems with the same value of $h/k_B T$ but different values of $h$ and $k_B T$. 
Values of $z$ and $d$ in Eq.\ \ref{eq:specpoft} are taken from fit values of $\alpha$, $\beta$ and the measured value of $\delta(h)$ as reported in Section \ref{sec:modelmethods}; note that these vary somewhat with $T$, giving different $p^{*}$ at the same $h/k_B T$.
Data agree excellently with predictions at short and long times.
The merely qualitative agreement at intermediate times is no cause for concern, but is an interesting `feature', because Eq.\ \ref{eq:pactiveoft} ignores all physics arising from the important fact that the sticky monomers are embedded, at a concentration $c_{st}$, in chains of length $N$, in a dense polymer melt.
The slower convergence of simulation results for $p_{active}(t)$ relative to the prediction of Eq.\ \ref{eq:specpoft} is consistent with such polymeric effects; better agreement is observed for dimer systems.  
As expected, the polymeric slowdown is greater at lower $T$.

The results above demonstrate the ability of our method to capture the effect of polymer physics  on nonequilibrium ``chemical dynamics''. 
Thus, as in Ref.\ \cite{combinedhuang01}, it can be used to perform ``$T$-jump'' simulations.
These may be useful in analyzing phenomena observed in recent real $T$-jump experiments; nonequilbrium sticky bond behavior is also expected to play a role in self healing AP systems \cite{cordier08}.
Note, for example, that the timescale over which $p_{active}$ changes in Figure \ref{fig:pactivefig} is smaller than the equilibrium $\tau_{sb}$ ($1.5\cdot10^{3}\tau_{LJ}$ for $h = 10u_0$).
Similarly, the timescale of self healing at a fractured surface (where $p_{active}$ is out of equilibrium) was found to be smaller than the time scale for near-equilibrium creep relaxation \cite{cordier08}.

\subsection{Nonlinear and Nonequilibrium Mechanical Properties}
\label{subsec:nonlinmechprop}

\begin{figure}[htbp]
\includegraphics[width=3.375in]{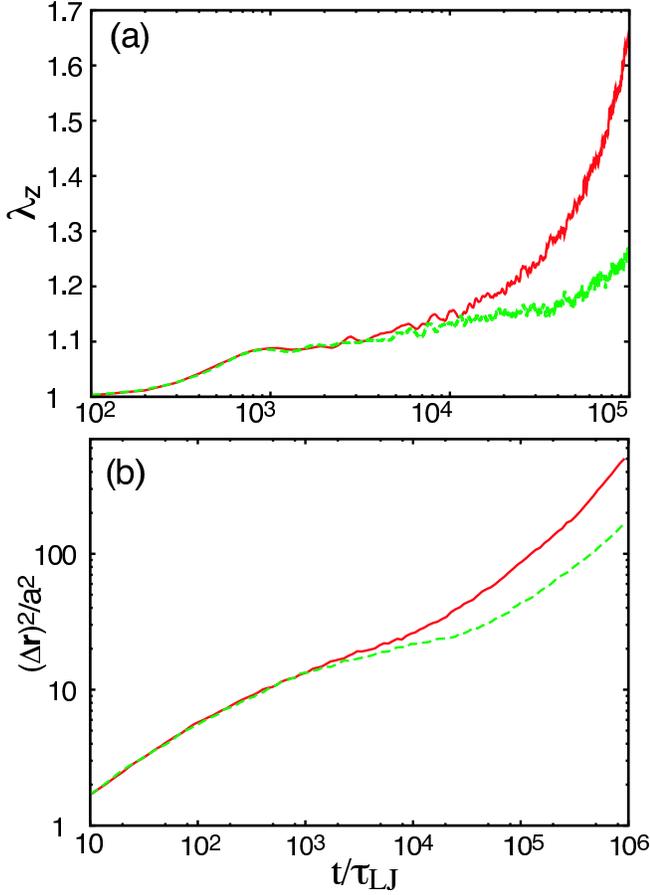}
\caption{Linear creep and quiescent diffusion for two systems with the same $\tau_{sb}$ but different SB recombination.  Results are for systems with $N_{ch}=5600$, $N=50$, and $k_B T = 1.0u_0$.  Panel (a) Stretch $\lambda_z = L_z/L_{z}^{0}$ under a creep stress $\Delta \sigma = .01u_0/a^{3}$.  The upper curve is for $h=10u_0$, $\tau_{MC}=10\tau_{LJ}$ and the lower curve is for $h = 12.5u_0$, $\tau_{MC} = 1.0\tau_{LJ}$.  Panel (b): mean squared displacement for the same systems in the \textit{quiescent} state.}
\label{fig:h10125creep01}
\end{figure}

Figure \ref{fig:h10125creep01}a shows results for creep tests of two systems with different thermodyamics and kinetics but the same $\tau_{sb}$ ($\tau_{sb} \simeq 1.5\cdot10^{4}\tau_{LJ}$).
Both tests were performed at $k_B T= 1.0u_0$.
The applied stress difference $|\sigma_{z} - (\sigma_x + \sigma_y)/2| = .01u_0/a^{3}$ is small.
At times $t \ll \tau_{sb}$, the extension ratio $\lambda_{z} = L_z/L_{z}^{0}$ is the same for both systems.
For $t \gg \tau_{sb}$, $\lambda_z$ is nearly linear in $ln(t)$, implying that the flow is nearly-linear creep.
The system with stronger bonds and greater SB recombination shows greater resistance to flow, i.\ e.\ a smaller creep compliance.

It is interesting to relate the creep response to the quiescent dynamics.
Figure \ref{fig:h10125creep01}b shows (quiescent) diffusion in the same systems.
The creep response and diffusion are remarkably similar; the onset of more rapid creep in the $h = 10u_0$, $\tau_{MC} = 10\tau_{LJ}$ system under stress corresponds directly to the onset of (relative) delocalization in the quiescent state.
This is consistent with a recent experiment showing connections between creep behavior and linear rheology in reversible supramolecular networks \cite{cordier08}.

Next we consider constant volume tension simulations at $h = 11.25u_0$, $k_B T = u_0$, and various $\tau_{MC}$.
These simulations can be considered to be an extension of Ref.\ \cite{rottach07}, which allowed breaking and formation of interchain bonds only at a few (discrete) strains; here SBs break and reform continuously.
Here we present results for $\dot{\epsilon} = 10^{-5.5}/\tau_{LJ}$.
Simulations at other $\dot{\epsilon}$ were considered; larger values $\dot{\epsilon} \gtrsim t_{cage}^{-1}$ make the non-SB-related viscous stress contribution unacceptably large, while smaller values lead to more sticky bond breaking/formation during deformation than is desirable at the values of $h$ and $\tau_{MC}$ considered.

\begin{figure}[htbp]
\includegraphics[width=3.375in]{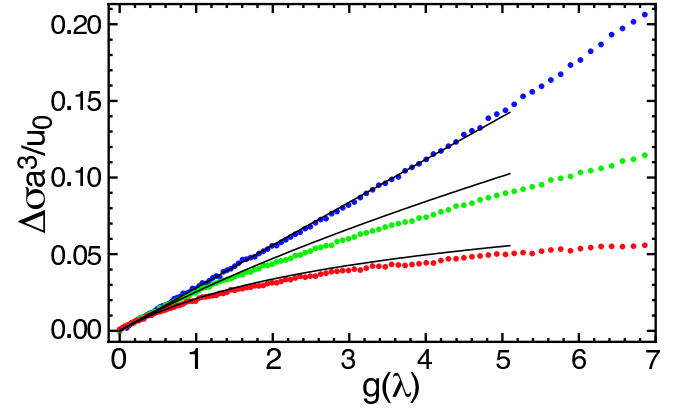}
\caption{Constant volume tension simulations for systems with different SB kinetics.  The stress difference $\delta \sigma = \sigma_z - (\sigma_x + \sigma_y)/2$ is plotted against $g(\lambda)=\lambda^{2}-1/\lambda$.  Systems have $N_{ch}=5600$, $N=50$, $h = 11.25u_0$, $k_BT = 1.0u_0$, and the strain rate is $\dot\epsilon \equiv \dot\lambda/\lambda = 10^{-5.5}/\tau_{LJ}$.  These runs use $\tau_0=.2\tau_{LJ}$ for greater accuracy.  Data from top to bottom correspond to $\tau_{MC}=\infty$ (no SB breaking/forming allowed during deformation), $\tau_{MC} = 100\tau_{LJ}$ and $\tau_{MC}=10^{1.5}\tau_{LJ}$. The solid lines are predictions of Eq.\ \ref{eq:simplestressmem}, with the value of $G_e$ taken from a fit to the $\tau_{MC}=\infty$ data and values of $\tau_{sb}$ taken from Table \ref{tab:htauMCtausb}.}
\label{fig:h1125constVtens}
\end{figure}

Figure \ref{fig:h1125constVtens} shows the stress difference $|\sigma_{z} - (\sigma_{x} + \sigma_{y})/2|$.
With MC deactivated during deformation ($f_{MC} = 0$ or equivalently $\tau_{MC}=\infty$), the stress takes a form close to that predicted by entropic elasticity \cite{treloar75}: $\sigma=G_e g(\lambda)$, where $g(\lambda) = \lambda^{2} - 1/\lambda$ and $\lambda = L_z/L_z^{0}$ as above.
$G_e$ is predicted to be $Nc_{st}p_{inter} k_BT/2$, where $p_{inter} < p^{*}$ is the interchain portion of active SBs; the actual value from the fit, $G_e = .028u_0/a^{3}$, is close to the predicted value $.031u_0/a^{3}$, indicating viscous stresses are low at this strain rate.
The fit is performed for $g \leq 5$; the nonlinear behavior observed at higher $g$ arises from finite extensibility of chain segments between crosslinks (as in standard nonlinear rubber elasticity \cite{treloar75, arruda93}).

The simplest result, assuming that $\dot\epsilon^{-1} \gg \tau_{free}$, $\tau_{sb} \gg \tau_{free}$ (here $\tau_{free}$ is the lifetime of unbonded SMs), and that SB breaking and formation rates do not vary with stress/strain, so that stress memory is lost like $exp(-t/\tau_{sb})$, is \cite{green46}
\begin{equation}
\begin{array}{l}
\sigma_{z}(\lambda)  = G_e exp(-ln(\lambda)/\dot\epsilon\tau_{sb})\times \\
 \\
\left[g(\lambda) + \displaystyle\frac{1 - \lambda^{2}}{1 - 2\dot\epsilon\tau_{sb}} + \displaystyle\frac{1/\lambda - 1}{1 + \dot\epsilon\tau_{sb}}\right] 
\end{array}
\label{eq:simplestressmem}
\end{equation}
where the first term in brackets is classical rubber elasticity and the second two terms reflect new SBs created during deformation.  
The $ln(\lambda)$ comes from the constant true strain rate $\dot\epsilon = \partial ln(\lambda)/\partial t$.

In Figure \ref{fig:h1125constVtens}, stress-strain results from simulations are compared to predictions from Eq.\ \ref{eq:simplestressmem} using values for $\tau_{sb}$ from Table \ref{tab:htauMCtausb} and the value of $G_e$ from the $\tau_{MC}=\infty$ system are shown.
We confine the comparison to the linear regime ($g \leq 5$) to avoid confusion.
Sticky bond recombination might be expected to slow relaxation.  
However, stress relaxation is actually faster than predicted by Eq.\ \ref{eq:simplestressmem}.
We have verified that $p_{active}$ does not decrease during deformation. 
It appears that instead, $\tau_{sb}$ is reduced by stress.
A detailed examination of this effect and comparison of nonlinear mechanical properties to theories, e.\ g.\  Ref.\ \cite{indei07} and transient network models, e.\ g.\ Refs.\ \cite{green46,tanaka92,vaccaro00},  is deferred to later work, but the data presented above suggests traditional theories will break down in the nonlinear regime.

\begin{table*}[htbp]
\caption{Variation of $\tau_{sb}$, $\tau^{*}$ and $D$ with $N$, $T$, $h$ and $\tau_{MC}$.  Times are in units of $\tau_{LJ}$.  Statistical errors are roughly $\pm 2\%$ or less.  All systems have $N_{ch} = 1400$ and data are averaged over multiple statistically independent states.  * denotes $\tau_0 = .5\tau_{LJ}$ results.  $--$ indicates calculation is prohibitive or we have insufficient data.   Results for $D$ in $N=1$ systems are not presented because they are negligibly affected by sticky bonding.}
\setlength{\extrarowheight}{-4pt}
\begin{ruledtabular}
\begin{tabular}{lcccccccc}
System & $N$ & $k_BT/u_0$ & $h/k_BT$ & $\tau_{MC}$ & $\tau_{sb}$ & $\tau^{*}$ & $\tau^{*}/\tau_{sb}$ & $10^{5}\tau_{LJ}D/a^{2}$\\
A & 50 & 1.0 &10 & 1 & $1.55\cdot10^{3}$ & $7.05\cdot10^{3}$ & 4.5 & 18.3\\
B & 50 & 1.0 &10 & 10 & $1.54\cdot10^{4}$ & $2.63\cdot10^{4}$ & 1.7 & 8.44\\
C & 50 & 1.0 &10 & 100 & $1.53\cdot10^{5}$ & $1.78\cdot10^{5}$ & 1.2 & 2.17\\
E* & 50 & 1.0 &11.25 & 0.5 & $2.40\cdot10^{3}$ & $2.28\cdot10^{4}$ & 9.5 & 9.31\\
DE* & 1 & 1.0 & 11.25 & $0.5^{*}$ & $3.35\cdot10^{3}$ & $2.23\cdot10^{4}$ & 6.7 & N/A\\
F & 50 & 1.0 &11.25 & 1 & $4.82\cdot10^{3}$ &  $2.95\cdot10^{4}$ & 6.1& 6.89\\
DF & 1 & 1.0 & 11.25 & 1 & $6.69\cdot10^{3}$ & $2.67\cdot10^{4}$ & 4.0 & N/A\\
G & 50 & 1.0 & 11.25 & 2 & $9.78\cdot10^{3}$ & $4.05\cdot10^{4}$ & 4.1 & 6.07\\
DG* & 1 & 1.0 & 11.25 & 2 & $1.34\cdot10^{4}$ & $3.46\cdot10^{4}$ & 2.6 & N/A\\
H & 50 & 1.0 &11.25 & 5 & $2.40\cdot10^{4}$ & $6.42\cdot10^{4}$ & 2.7 & 4.81\\
DH* & 1 & 1.0 & 11.25 & 5 & $3.36\cdot10^{4}$ & $5.50\cdot10^{4}$ & 1.6 & N/A\\
I & 50 & 1.0 &11.25 & 10 & $4.73\cdot10^{4}$ & $9.70\cdot10^{4}$ &  2.1 & 3.11\\
DI* & 1 & 1.0 & 11.25 & 10 & $6.42\cdot10^{4}$ & $8.52\cdot10^{4}$ & 1.3 & N/A\\
J & 50 & 1.0 &11.25 & $10^{1.5}$ & $1.55\cdot10^5$ & $2.32\cdot10^{5}$ & 1.5 & 1.95\\
DJ* & 1 & 1.0 & 11.25 & $10^{1.5}$ & $2.05\cdot10^5$ & $2.30\cdot10^{5}$ & 1.1 & N/A\\
K & 50 & 1.0 &11.25 & 100 & $4.82\cdot10^{5}$ & $6.28\cdot10^{5}$ & 1.3 & 0.94\\
DK* & 1 & 1.0 & 11.25 & 100 & $6.70\cdot10^{5}$ & $6.85\cdot10^{5}$ & $\sim1.02$ & N/A\\
L & 50 & 0.6 & 11.25 & 1 & $1.88\cdot10^{3}$ & $8.02\cdot10^{3}$ & 4.3 & $--$\\
M & 50 & 0.6 & 11.25 & 10 & $1.91\cdot10^{4}$& $3.15\cdot10^{4}$ & 1.6 & $--$\\
N & 50 & 0.6 & 11.25 & 100 & $1.93\cdot10^{5}$ & $2.22\cdot10^{5}$ & 1.2 & $--$\\
O & 50 & 1.0 &12.5 & 1 & $1.53\cdot10^{4}$ & $1.23\cdot10^{5}$ & 8.0 & 2.79\\
P & 50 & 1.0 &12.5 & 10 & $1.55\cdot10^{5}$ & $3.58\cdot10^{5}$ & 2.3 & $--$\\
Q & 50 & 1.0 & 12.5 & 100 & $1.53\cdot10^{6}$ & $2.07\cdot10^6$ & 1.3 & $--$\\
R & 50 & 1.0 & 13.75 & 1 & $4.90\cdot10^{4}$ & $5.42\cdot10^{5}$ & 11 & $--$\\
S & 50 & 1.0 & 13.75 & 100 & $4.62\cdot10^{6}$ & $7.34\cdot10^{6}$ & 1.6 & $--$\\
T & 50 & 1.0 & 15 & 1 & $1.56\cdot10^{5}$ & $2.49\cdot10^{6}$ & 16 & $--$\\
U & 50 & 1.0 & 15 & 100 & $1.48\cdot10^{7}$ & $2.53\cdot10^{7}$ & 1.7 &  $--$\\
V & 50 & 1.0 & 16.25 & 1 & $5.0\cdot10^{5}$ & $1.1\cdot10^{7}$ & 22 & $--$\\
\end{tabular}
\end{ruledtabular}
\label{tab:htauMCtausb}
\end{table*}

\section{Discussion and Conclusions}
\label{sec:conclude}

We have performed an initial set of simulations using a new coarse-grained model for associating polymers.
The MD/MC hybrid algorithm and variable chemical kinetics allow for greater realism and flexibility than in previous simulations of AP networks.
Further, the 1-1 sticky monomer binding topology imposed here reflects current experimental trends.
The model was extensively validated and is able to accurately model equilibrium dynamical properties, nonlinear mechanical properties, and far-from-equilibrium systems.
We studied the model over a very broad parameter space.
While have emphasized that we study APs by analogy because simulations of chemically realistic AP networks are not yet computationally feasible \cite{foot8}, our results should nevertheless aid in ``rational'' \cite{loveless05} design of AP systems, especially in ``transition'' regimes like those discussed in this paper.

The key results presented here focused on separation, comparison and contrast of thermodynamic and chemical-kinetic effects on SB recombination, the motion of individual chains, and bulk mechanical properties.
As expected, instantaneous network structure was independent of kinetics at fixed thermodynamic conditions (i.\ e.\ sticky bond strength $h/k_B T$), and relaxation times increases with increasing $h/k_B T$.
Similarly, at fixed $h$ and $k_B T$, relaxation slows as the chemical kinetics are slowed.
This was illustrated by measurements of monomer diffusion.
In the physical gel regime, monomers experience a temporary ``caging'' similar to that found in glasses.
This caging effect strengthens as SB strength is increased \cite{kumar01,baljon07} but also as kinetics are slowed at fixed SB strength.
Analyses showed that chains become increasingly localized in a manner similar to that associated with the increase in dynamical heterogeneity in non-AP melts approaching the glass transition.
Of course, the analogy should not be taken too far; in AP networks the caging is produced only by sticky monomers while in systems approaching $T_g$ it is produced by hard core repulsions of all monomers.

We find, as expected, that the chemical kinetics controlling $\tau_{sb}$ are ``mean-field'' \cite{cates87} and mappable to a two-state Arrhenius model.
However, as kinetic rates are increased and SB recombination becomes non-kinetically-limited, the relation between the SB lifetime $\tau_{sb}$ and other relaxation times, such as the effective SB lifetime $\tau^{*}$, becomes decidedly nontrivial.
This was explicitly related to the crossover to diffusion limited SB recombination.
A new quantitative relation between $\tau^{*}$ and $\tau_{sb}$ was found.
Such relations should be of interest because rheological experiments can typically only access $\tau^{*}$, which is assumed to control stress relaxation (e.\ g., because scission followed by quick recombination does not relax stress) \cite{cates87, rotello08, katieunpublished}.

While the results for $T$-variation and mechanical properties presented here were limited and somewhat preliminary, we showed examples which illustrate important effects.
Analysis of diffusion on intermediate time scales illustrated the point that sticky bond and underlying polymeric timescales will in general vary differently with $T$, affecting the `interplay' in nontrivial ways.  
In two systems with the same $\tau_{sb}$, systems with different propensities for SB recombination showed the same creep flow at short times, but those with greater recombination showed a smaller long-time (``DC'') creep compliance.
These differences were directly related to the faster delocalization of chains in the \textit{quiescent} state for systems with less recombination.
Constant-volume deformation studies showed that, as expected, $\tau_{sb}$ is reduced by stress.
Extensive studies of the variation with $T$ (in systems including attractive nonbond interactions for greater realism) and more detailed analyses of nonlinear mechanical properties are underway.

Nearly all published analytic theories for AP networks assume a single controlling relaxation time, either $\tau_{sb}$ or $\tau^{*}$, controls the ultimate relaxation properties (i.\ e.\ other relaxation times scale with the controlling time).
We showed though various measurements that there is a broad parameter space  (both in terms of SB strength and kinetics) within the physical gel regime where the ``scaling'' assumption fails.
This parameter space corresponds to the conditions (1) $\tau^{*}/\tau_{sb}$ is larger but not ``much larger'' than unity, and/or (2) SB recombination is not kinetically limited.
Deviations from this ``scaling'' behavior due to multiple controlling relaxation times have been observed \cite{loveless05,xu09}, but had not yet been well understood.
These deviations had been previously assumed to arise from chemical disorder, and this is no doubt partially correct, but as discussed in this paper, they also arise from the `interplay' between SB thermodynamics, kinetics, and polymer physics.
If either (1) or (2) hold, both traditional \cite{tanaka92,vaccaro00} and more sophisticated \cite{leibler91,rubinstein98,rubinstein01} theories should fail to predict the mechanical properties of AP networks.
This is not meant as a criticism of the theories, merely an observation that there is a broad parameter space where one or more of their assumptions fail.

The DL and KL limits have been discussed by O'Shaughnessy and Yu \cite{oshaugh95}; they respectively correspond to dominance of the $K$-term and $C$-term in our Eq.\ (\ref{eq:kltlpowerlaw}).
Conditions under which systems may lie outside the KL limit and/or evidence for systems which lie outside it are also discussed, to some extent, in the context of AP networks in Refs. \cite{loverde05, manassero05, yount05}.  
Coupling between SB and polymeric relaxation has also been treated approximately by Cates \cite{cates87} and Leibler \textit{et.\ al.} \cite{leibler91}, respectively for linear EP systems and AP networks where recombination is improbable.
Among published analytic theories for AP networks, Refs.\ \cite{leibler91,oshaugh98} qualitatively treat non-kinetically-limited systems and Ref.\ \cite{rubinstein98} treats SB recombination.
Our results are consistent with the argument of Ref.\ \cite{oshaugh98} that reaction rates (here defined as non-recombinative SB exchange) reach the mean-field/KL regime only when reaction is slow compared to the longest ``underlying'' polymeric relaxation time [in our case $min(t_{cage},t_{chem})$] \cite{foot13}.
Interestingly, Refs.\ \cite{degennes82,oshaugh88} suggest that MF kinetics would apply to $\tau^{*}$ as well as $\tau_{sb}$ in dimensions $d \geq 4$ because $<|\delta \vec{r}|^{2}> \sim t^{y}$ necesarily has $y \leq 4$ (i.\ e.\ no diffusion-limited regime is possible).  
This suggests that diffusion-limited SB recombination, which increases $\tau^{*}/\tau_{sb}$, will increase in importance in AP systems with effectively reduced dimensionality (e.\ g.\ very thin films or ``pores'').  
Combining the approaches of Refs.\ \cite{leibler91, oshaugh98, rubinstein98} may be useful for developing optimal analytic theories of these systems, at least for $T$ well above $T_g$.
However, we are not aware of any quantitative discussion of the crossover between the DL and KL regimes such as presented here \cite{foot14}.

The rheologically simple (i.\ e.\ all key relaxation times scale with $\tau_{sb}$ \cite{yount05,loveless05}) behavior observed in the majority of experiments on AP networks indicates they exhibit KL behavior.
Note that these experiments have shown KL behavior even though their values of $p^{*}$ are comparable to values for systems which in our model exhibit DL behavior at low $\tau_{MC}$.
This likely arises from the slow kinetics caused by the bulkiness and directional interactions of real sticky monomers.
Creating (real) strong-binding SMs with even faster kinetics seems to be difficult.
However, one can move out of the KL regime (at fixed kinetic rates) simply by slowing the polymeric relaxation times, e.\ g.\ by going to higher concentrations and/or entangled chains.
Experiments in this regime are underway \cite{xu09}, and seem to show a breakdown of the simple scaling; for example, they show an unusually high power law dependence of viscosity on concentration, which appears to arise because $\tau_{sb}$ is of order the time scale for reptation.
Ref.\ \cite{loveless05} also shows an apparent (if weak) breakdown in scaling at the highest frequencies considered.  
In the context of these observations, we note that the parameter space where (1) and or (2) hold may be of greatest interest for designing materials with novel mechanical properties.
Our model seems well suited to aid in understanding the complicated behavior of AP networks in this regime.

Here we have left the regime of physically entangled APs, which is the regime treated by some key analytic AP theories \cite{leibler91,rubinstein01,semenov06}, untouched.
Also, the ``interplay'' described in this paper should depend on the details of sticky monomer arrangement along chains, not just $N$ and $c_{st}$.
Studies of systems with a wide range of $N$ and $c_{st}$, as well as inhomogeneous (chemically disordered) systems, are underway.

In real AP networks the sticky and regular monomers have different sizes and chemistries.
Thus an obvious extension of our model would be to increase the differences between the sticky and normal monomers.  
For example, changing secondary interactions may induce microphase separation \cite{manassero05,grest96}.
Another extension would be use of a more realistic sticky bonding potential such as those used to model H-bonds (see e.\ g.\  Ref.\ \cite{karimi08} and refs.\ therein), but here we have focused on chemistry-independent properties.
A coarse-grained way to to capture this would be to keep the binary bonding rules, but include more than one sticky site per SM; this would increase the directionality of bonding, which is a key to the performance of the UPy systems \cite{sijbesma97}.
Finally, nanocomposites of associating polymers \cite{sprakel07}, where the presence of nanoparticles may or may not \cite{sen07,allegra08} affect single-chain structure but will certainly affect AP network structure, should have even richer physics than regular polymer nanocomposites.

Michael Rubinstein, Kathleen E.\ Feldman, Arlette R.\ C.\ Baljon, Phillip A. Pincus, Edward J.\ Kramer, Frank L.\ H.\ Brown, Stephen L.\ Craig, Gary S. Grest, Richard C. Elliott, Stuart J.\ Rowan, Steven J. Plimpton, and Ronald G. Larson provided helpful discussions.
Michael Rubinstein additionally provided helpful comments on the manuscript.
Brian B.\ Rochford assisted with the numerical analysis of $P_{recomb}$.
This work was supported by the MRSEC Program of the National Science Foundation under Award No.\ DMR05-20415.

\begin{appendix}

\section{Percolation Gel Transition}
\label{app:percgt}

Let $N^{cl}_{i} \equiv N_{ch}P(N_{ch},i)$ be the average number of disconnected clusters of $i$ chains in a system of $N_{ch}$ chains with periodic boundary conditions (at any given time).
$P(N_{ch},i)$ is a cluster size probability distribution with $\sum_{i=1}^{N_{ch}} iP(N_{ch},i) \equiv 1$.
The number averaged cluster size is then $N_n = N_{ch}\sum_{i=1}^{N_{ch}} iP(N_{ch},i)$ and the weight averaged cluster size is $N_w = N_{n}^{-1}N_{ch}\sum_{i=1}^{N_{ch}} i^{2}P(N_{ch},i)$.
For an infinite system,  the percolation gel transition occurs (by definition) when $p^{*}$ exceeds $p_{perc}$; $N_w$ diverges at $p^{*}=p_{perc}$ \cite{rubinstein03}. 
However, computer simulations are limited to finite $N_{ch}$, and the value of $N_w$ cannot exceed $N_{ch}$.
Thus the $N_{ch}$-dependent geometric percolation $p^{*} = p_{span}$ (at which one aggegrate spans the system) approaches $p_{perc}$ from below as $N_{ch}\to\infty$) \cite{combinedgrest90}.
Fortunately, $p_{perc}$ and $h_{perc}$ (the value of $h$ at which $p^{*}=p_{perc}$ in an infinite system) can be estimated for finite $N_{ch}$ using a standard finite size analysis \cite{frenkel02}.

\begin{figure}[h]
\includegraphics[width=3.375in]{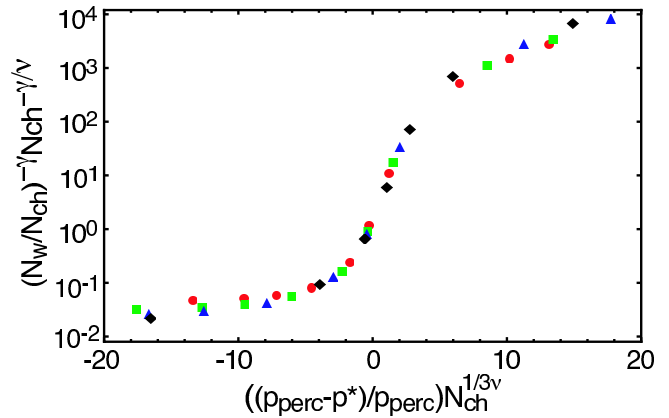}
\caption{Finite-size analysis of the percolation gel transition.  Data are at $k_B T = 1.0u_0$ for $N_{ch} = 700$ (circles), $N_{ch} = 1400$ (squares), $N_{ch} = 2800$ (triangles), and $N_{ch} = 5600$ (diamonds) are shown. $p_{perc}\simeq .40$.}
\label{fig:PercFSanalys}
\end{figure}

We perform such an analysis, following Ref.\ \cite{pandey85}.
Figure \ref{fig:PercFSanalys} shows this analysis for $N=50$, $c_{st}=.08$ systems at $k_B T = 1.0u_0$.
The figure plots the rescaled variables $(N_w/N_{ch})^{-\gamma}N_{ch}^{-\gamma/3\nu}$ vs.\ $((p_{perc} - p^{*})/p^{*})N_{ch}^{1/3\nu}$ \cite{pandey85}.  
$p^{perc} \simeq .40$ is close to the predicted value $1/(Nc_{st}p^{*}f_{inter}-1)$ \cite{combinedgrest90}, where $f_{inter}$ is the fraction of SBs which are interchain rather than intrachain.
The exponents used to collapse the data in the figure are $\gamma \simeq 1.7$ and $\nu \simeq 1.2$; considering a narrower range of $h$ gives values consistent with predictions from the theory of critical phenomena ($\gamma \simeq 1.8,\ \nu \simeq 0.9$) \cite{degennes79}. 
These exponents been extensively discussed in the literature \cite{pandey85,combinedgrest90,liu96} and need not be discussed further here.
In the figure, $h$ increases going from right to left, and percolation occurs at $h_{perc} \simeq 4.25u_0$.

\section{Numerical Analysis of SB Recombination}
\label{app:SBrecomb}

In this paper, the sticky bond self-correlation function $P_{auto}(\triangle t)$ is the probability that a bond between two given SMs exists both at times $t$ and $t + \triangle t$, while the SB ``transition function'' $P_{trans}(\triangle t)$ is the probability that the bond exists \textit{continuously} between times $t$ and $t + \triangle t$. 
The SB ``recombination function'' $P_{recomb}(\triangle t) \equiv P_{auto}(\triangle t) - P_{trans}(\triangle t)$ is the probability that a pair of SMs will be bonded at two times separated by $\triangle t$ but that the bond between them has broken at least once during that interval.
We find $P_{trans}$ exhibits nearly single-exponential decay, $P_{trans} = \exp(-t/\tau_{sb})$ for all systems, while for $h \gtrsim 10u_0$, $P_{auto}$ also shows exponential decay for (at least) the first decade.
Values of $\tau^{*}$ presented here are measured from fits to  $P_{trans} = \exp(-t/\tau^{*})$.
All quantities are averaged over all SM pairs and all $t$.

\end{appendix}


\begin{thebibliography}{117}
\bibitem[{foo({\natexlab{a}})}]{foot7}
\bibinfo{note}{It is impossible to cite all such studies; there have been very
  many. We focus on those most relevant to the current work.}
\bibitem[{foo({\natexlab{b}})}]{foot9}
\bibinfo{note}{Notable exceptions are Refs.\ \cite{baxandall89,semenov06}.}
\bibitem[{cra()}]{craigexpl}
\bibinfo{note}{These experiments were largely performed with `floating' sticky
  monomers which can attach to any monomer on the parent chains, but since the SMs are
  bonded to at least one chain the vast majority of the time, for practical
  purposes the systems are APs as defined here.}
\bibitem[{foo({\natexlab{c}})}]{foot1}
\bibinfo{note}{For the NPT and deformation simulations we use a smaller
  timestep $\delta t = .008\tau_{LJ}$. For all simulations, because of the weak
  thermostatting and large timestep, we set a cap $0.075a\delta t$ on the
  maximum monomer displacement per timestep, to avoid crashes at large $h$.
  This restriction is rarely actively applied (the associated kinetic energy is
  $\sim 50k_B T$) and the need for it can be eliminated by using smaller
  $\delta t$}.
\bibitem[{foo({\natexlab{d}})}]{foot4}
\bibinfo{note}{The difference between $U_{sb}(r,h)$ and the potential used in
  Ref.\ \cite{baljon07} is that they absorbed the last two terms into a single
  offset $U_{assoc}$. The value $U_{assoc} = 22u_0$ used in Ref.\
  \cite{baljon07} corresponds to $h=4.1715u_0$.}
\bibitem[{foo({\natexlab{e}})}]{foot8}
\bibinfo{note}{Except perhaps at high (processing) temperatures; for example,
  the variation of $k_b$ with $T$ for UPy in $CDCl_{3}$ \cite{sontjens00}
  indicates $h/k_{B}T \simeq 25$ for $T \simeq 300K$, so the same system at $T
  = 450K$ would fall within our studied range of $h/k_B T$.}
\bibitem[{foo({\natexlab{f}})}]{foot15}
\bibinfo{note}{A key difference is that the ``cage'' is typically much larger
  in an AP network well above $T_g$ then it is in the same system near and
  below $T_g$; it is produced only by the sticky monomers in the former, but by
  all monomers in the latter.}
\bibitem[{foo({\natexlab{g}})}]{foot6}
\bibinfo{note}{For $\tau_{MC}=\tau_{LJ}$. This value could in principle change
  as kinetics are slowed.}
\bibitem[{foo({\natexlab{i}})}]{foot17}
\bibinfo{note}{This is valid for dense melts.  If chains overlap less strongly, the dependence of $a_{cage}$ on $c_{st}$ can be different; see Ref.\ \cite{rubinstein98}.}
\bibitem[{foo({\natexlab{j}})}]{foot12}
\bibinfo{note}{Notable exceptions are Refs.\
  \cite{degennes82,fredrickson96,oshaugh98}, but these considered the simpler
  case of irreversible SB formation. It would be interesting to see if these
  theories could be generalized to reversible AP networks.}
\bibitem[{foo({\natexlab{k}})}]{foot13}
\bibinfo{note}{Note that given that bond lifetimes are exponentially
  distributed rather than single-valued, it is not enough that $\tau_{sb} >
  min(t_{cage},t_{chem})$; it must be MUCH larger to reach the KL regime.}
\end{thebibliography}

\end{document}